\documentclass[aps,pre, twocolumn, groupedaddress, 10pt]{revtex4-1}
\usepackage{amsmath}

\usepackage{tabularx}
\usepackage{bbding} 
\usepackage{amssymb}
\usepackage{amsfonts}
\usepackage[framemethod=TikZ]{mdframed}
\usepackage{xcolor} 
\usepackage{setspace}
\usepackage{titlesec} 
\usepackage{caption}
\usepackage[T1]{fontenc} 
\usepackage{subcaption}
\usepackage{lipsum}
\usepackage{tcolorbox} 
\usepackage{mathtools} 
\usepackage{siunitx}
\usepackage{booktabs}
\usepackage{xr} 

\allowdisplaybreaks

\usepackage{enumerate}

\usepackage[hidelinks]{hyperref} 

\DeclareMathOperator{\Tr}{Tr}
\DeclareSymbolFont{rmlargesymbols}{OMX}{mdbch}{m}{n}
\DeclareMathSymbol{\rmintop}{\mathop}{rmlargesymbols}{82}

\DeclareMathSymbol{\rmointop}{\mathop}{rmlargesymbols}{72}

\newcommand{\refew}[1]{Eq.\eqref{eq:#1}}
\newcommand{\rfw}[1]{Eq.\eqref{eq:#1}}
\newcommand{\reffig}[1]{Fig. \ref{fig:#1}}
\newcommand{\refsec}[1]{Sec. \ref{sec:#1}}

\newcommand{\mm}[0]{\nonumber \\}

\newcommand{\tbf}[1]{\textbf{#1}}

\newcommand{\TII}[0]{T_{\text{II}}}
\newcommand{\TI}[0]{T_{\text{I}}}

\def\@caption@fignum@sep{. }

\allowdisplaybreaks

\begin{document}


\title{Self-assembly of a dimer system}



\author{Mobolaji Williams} 
\affiliation{%
Department of Physics, Harvard University, Cambridge, MA 02138, USA 
}%
\email{mwilliams@physics.harvard.edu}


\date{April 22, 2019; \tbf{updated:} April 26, 2021}

\begin{abstract}

In the self-assembly process which drives the formation of cellular membranes, micelles, and capsids, a collection of separated subunits spontaneously binds together to form functional and more ordered structures. In this work, we study the statistical physics of self-assembly in a simpler scenario: the formation of dimers from a system of monomers. The properties of the model allow us to frame the microstate counting as a combinatorial problem whose solution leads to an exact partition function. From the associated equilibrium conditions, we find that such dimer systems come in two types: "search-limited" and "combinatorics-limited", only the former of which has states where partial assembly can be dominated by correct contacts. Using estimates of biophysical quantities in systems of single-stranded DNA dimerization, transcription factor and DNA interactions, and protein-protein interactions, we find that all of these systems appear to be of the search-limited type, i.e., their fully correct dimerization regimes are more limited by the ability of monomers to find one another in the constituent volume than by the combinatorial disadvantage of correct dimers. We derive the parameter requirements for fully correct dimerization and find that rather than the ratio of particle number and volume (number density) being the relevant quantity, it is the product of particle diversity and volume that is constrained. Ultimately, this work contributes to an understanding of self-assembly by using the simple case of a system of dimers to analytically study the combinatorics of assembly. 

\end{abstract}


\maketitle


\section{Introduction}

Self-assembly occurs in many microbiological systems, driving the formation of bilayer membranes, micelles, and virus capsids \cite{nelson2004biologicalch8}. For a macromolecular system to be able to undergo self-assembly, its components must be able to find one another within their larger volume and also be able to distinguish correct from incorrect contacts. In the process of the system evolving towards its final configuration, the number of possible incorrect contacts is always much greater than the number of correct contacts, a fact which makes the  mathematical problem of self-assembly a combinatorial one. 

As a brute force resolution to this combinatorial problem, researchers have often used computational methods to study the specific properties of self-assembled systems \cite{wales2005energy,nguyen2007deciphering,johnston2010modelling}. Conversely, analytical studies of self-assembly often avoid combinatorics all together and begin under the infinite volume-infinite particle number assumptions of the law of mass action \cite{israelachvili1976theory, marsh2012thermodynamics, perlmutter2015mechanisms} or, in order to avoid the complications associated with analyzing a specific system, have focused on more phenomenological properties of self-assembly \cite{nguyen2016design, norn2016computational}. 

\begin{figure*}[t]
\begin{centering}
\begin{subfigure}{0.3\textwidth}
\centering
\includegraphics[width=\linewidth]{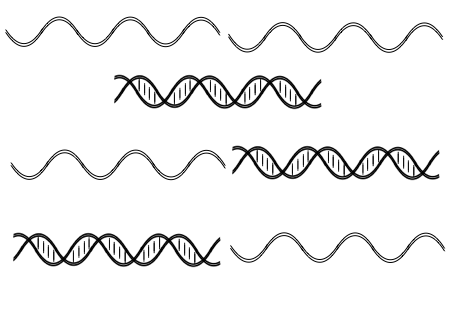}
	\caption{}
	\label{fig:ssdna}
\end{subfigure} \hfill
\begin{subfigure}{0.3\textwidth}
\centering
\includegraphics[width=\linewidth]{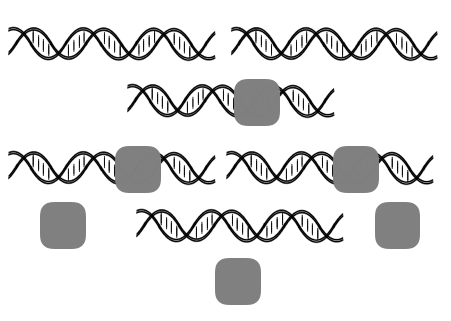}
	\caption{}
	\label{fig:protein_dna}
\end{subfigure} \hfill
\begin{subfigure}{0.3\textwidth}
\centering
\includegraphics[width=\linewidth]{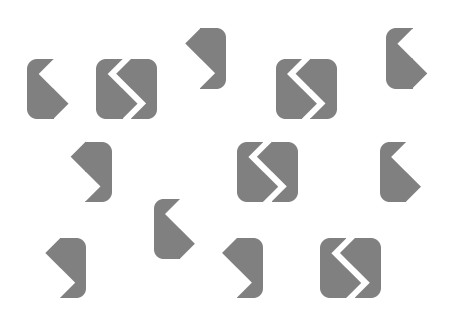}
	\caption{}
	\label{fig:protein_protein}
\end{subfigure} 
	\caption{Self-assembling biomolecular dimer systems. In (a), distinct single-stranded DNA (ssDNA) strands exist in a system with their complementary strands and with other double-stranded DNA (dsDNA). In (b), transcription factors (TFs) exist in a system with their binding sites on DNA and with already bound TF-DNA dimers. Since the binding sites are embedded in the much longer strand of an entire DNA molecule, the effective DNA molecules to which the TFs bind are much less motile than the TFs. In (c), distinct protein monomers exist in a system with the heterodimers formed from them. In all systems, we consider "fully correct assembly" or "fully correct dimerization" as the state where all monomers are bound to their correct monomer or binding site.}
	\label{fig:dimer_int}
\end{centering}
\end{figure*}

However, it is possible to study self-assembly analytically and specifically in the context of a model whose combinatorial properties are simple enough to admit an exact expression for the partition function. Although the typical examples of self-assembly involve the creation of large macromolecular structures on time scales relevant for cellular function, a simple kind of self-assembly is exemplified in the way single-stranded DNA fragments attach to their complementary strands, transcription factors find their correct DNA binding sites, and proteins seek their optimal binding partners (\reffig{dimer_int}).  In all of these systems, as in all systems capable of self-assembly, monomers only assemble into a functional set of interactions if the monomers can find one another and bind correctly. 

We can capture the basic features of these systems with a simple model. Say we have $2N$ distinct monomers $\alpha_1, \alpha_2, \ldots, \alpha_{2N}$ which form correct or incorrect contacts with one another according to the reaction equation
\begin{equation}
\alpha_i + \alpha_j
\xrightleftharpoons[ ]{} \alpha_i \alpha_j.
\label{eq:rxnrate}
\end{equation}
With $2N$ monomers, there are $N(2N-1)$ possible $(\alpha_{k}, \alpha_{\ell})$ pairs, and we define $N$ of these pairs as "correct" contacts that have a lower binding energy than that of the remaining $2N(N-1)$ contacts which are labeled as "incorrect". The binding energy is $-(E_0+\Delta)$ for correct contacts compared to $-E_0$ for incorrect ones, where $E_0, \Delta>0$. We say the system has undergone "fully correct dimerization" when all monomers are bound to their correct partners. 

In spite of the apparent simplicity of this model, the correct and incorrect interactions are defined by non-trivial combinatorics which lead to a unique partition function and surprising phase behavior of the self-assembled system. In particular, for a system of monomers contained in a volume $V$ and satisfying $N\gg1$, we find that the two necessary (but not sufficient) conditions the system must satisfy in order to be capable of fully correct dimerization are 
\begin{equation}
2N < e^{\beta \Delta}, \qquad NV < \sqrt{2}\, \lambda_0^3 \, e^{\beta(E_0+\Delta)},
\label{eq:conds_sa}
\end{equation}
where $\beta = 1/k_BT$ and $\lambda_0$ is the de Broglie thermal wavelength of a monomer. The first condition in \rfw{conds_sa} ensures that the energy advantage for correct contacts can overcome the combinatorial disadvantage of correct contacts. The second condition ensures that the monomers are able to find one another in their volume and bind together. What is interesting about these dual conditions is that, although one might think that number density is a relevant quantity in defining the possibility of self-assembly, the ratio of $N$ and $V$ does not appear, and instead it is their product and $N$ alone which are constrained. Moreover, both conditions in \rfw{conds_sa} can clearly only be satisfied under finite number and finite volume assumptions and thus a precise statistical physics formulation is required to obtain them. 

This problem of building models of correct and incorrect dimers has a few antecedents in the study of protein interactions. The authors of \cite{deeds2007robust} computationally studied the diffusion of dimer-forming lattice proteins in a three-dimensional grid and inferred that  low-energy specific dimers dominate higher-energy non-specific dimers, only if the system temperature is low enough that specific dimers are stable but high enough that non-specific dimers are unstable. The authors of \cite{zhang2008constraints} used the law of mass action \cite{sethna2006statistical_ch6} to study specific and non-specific protein interactions and establish approximate bounds on the minimum protein concentration and maximum protein diversity a cell requires to be in a safe zone, i.e., a parameter regime where non-functional interactions comprise fewer than 50$\%$ of the total interactions. In \cite{johnson2011nonspecific}, the authors employed a computational evolutionary model of protein interactions to show how selection pressure that seeks to minimize non-specific interactions can determine the way the energy gap between specific and non-specific interactions depends on the number of protein interfaces. 

What distinguishes the current work from these previous approaches is that it begins with simple assumptions concerning how correct and incorrect dimers can form from monomers (that are not necessarily proteins) and embeds these assumptions in an analytical statistical mechanics framework. Using such a framework allows us to both respect the finite-number properties key to defining the combinatorics of the system and to derive general equations governing dimer assembly rather than having to infer such equations from computational trends. 

The purpose of this work is to use statistical physics to better understand the properties of dimer self-assembly. In \refsec{ngpt}, we present the premises of our model, connect these premises to a combinatorial problem we name the "Dance Hall Problem," and then use the solution of this problem to compute the partition function of the system. In \refsec{eqbm}, we approximate the partition function through Laplace's method and obtain the equilibrium conditions relating the number of correct dimers to the total number of dimers in the system. In \refsec{phases}, we define the condition under which the dimer system undergoes fully correct dimerization, and use this condition to categorize dimer systems as one of two approximate types. In this section, we also numerically solve and plot the equilibrium conditions, compare the results to simulations, and depict the dimer system in parameter space. In \refsec{ineq}, we derive the necessary conditions for the system to be capable of fully correct dimerization, and interpret the two types as corresponding to "search" or "combinatorics" limits on fully correct dimerization. In \refsec{bio}, we apply the derived results to biomolecular systems of ssDNA-ssDNA interactions, TF-DNA interactions, and protein-protein interactions ultimately finding that all such systems appear to be of the search-limited type. In the next two sections, we outline ways to interpret the model, its limitations and how to extend it to better reflect the properties of real dimer systems. In \refsec{conclusion}, we summarize the conclusions of the paper, and in \refsec{supp}, we link to the code used to generate all figures and data tables.

\section{Non-Gendered Partition Function \label{sec:ngpt}}
In this section, we build the partition function for a system of distinguishable monomers that can form incorrect or correct dimers contingent on the dimer's constituent monomers. To match the physical conditions of self-assembly, we impose that the binding energy for the correct dimer is lower than the binding energy of the incorrect dimer, and thus that correct dimers are energetically preferred. However, the combinatorics of the dimer assembly is such that there are many more incorrect dimer microstates than correct dimer microstates, and so incorrect dimers are entropically preferred. We refer to this as the "combinatorial disadvantage" of correct dimers. 

We complete the calculation in steps: After outlining the particle and energy properties of the model, we present the partition function, reframe its computation in terms of a combinatorial sub-problem, and finally use the solution to this sub-problem to obtain an exact integral expression for the partition function.  

The system studied in this section (and presented throughout the main body of the paper) is termed "non-gendered" to emphasize the fact that  there is only one type of monomer and each monomer can bind to any other monomer. Such systems well describe the conditions of ssDNA-ssDNA interactions and some protein-protein interactions. But in TF-DNA interactions, there are two types of "monomers" each of which only binds to the other type; we call this system "gendered." In the Appendix, we outline the mathematical and physical properties of this "gendered" dimer model. 

\subsection{Naive Partition Function}

Say that our system contains $2N$ distinguishable monomers labeled $\alpha_1, \alpha_2, \ldots, \alpha_{2N}$. Each monomer has a mass $m_0$, and the monomers exist at thermal equilibrium temperature $T$ in a volume $V$.  Each monomer can bind to any other monomer, and when monomer $\alpha_k$ binds to monomer $\alpha_{\ell}$, the two form the dimer $(\alpha_k, \alpha_\ell)$ where the ordering within the pair is not important. 

\begin{figure}[t]
\centering
\includegraphics[width=.95\linewidth]{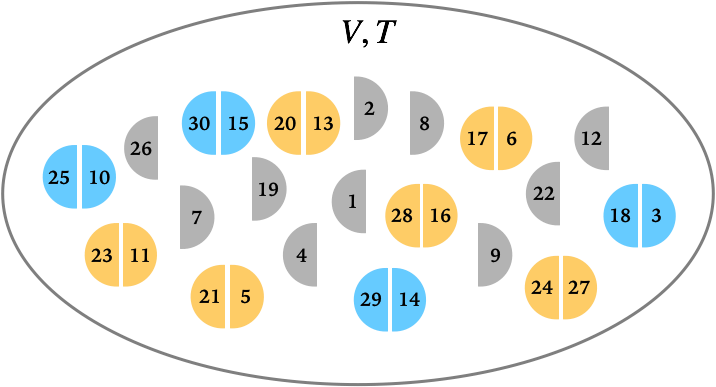}
	\caption{(Color online) Example microstate of the non-gendered system with $2N=30$ monomers. Correct dimers consist of binding $k$ to $k+15$ and have binding energy $-(E_0+\Delta)$. All other dimers are incorrect and have binding energy $-E_0$. This microstate has four correct dimers (in blue), six incorrect dimers (in yellow), and ten monomers (in grey).  For pictorial clarity, the figure represents monomers as half-circles, but monomers are taken to be point particles in the model. To which half-circle the individual monomers correspond is not important. The total binding energy for this microstate is $-(10 E_0 + 4 \Delta)$. }
	\label{fig:non_gen_sys}
\end{figure}

Without loss of generality, we define correct dimers as those consisting of an $\alpha_{k}$ binding with $\alpha_{N+k}$ where $k \leq N$; all other dimers are considered incorrect. Thus each monomer has one other monomer to which it binds to yield a correct dimer, and, more generally, there are $N$ possible correct dimers and $2N(2N-1)/2 - N = 2N(N-1)$ possible incorrect dimers. We take the incorrect dimers to form with binding energy $-E_0$, and the correct dimers to form with binding energy $-(E_0 + \Delta)$ where $E_0, \Delta>0$. Summarily, the binding energy for a dimer $(\alpha_i, \alpha_{j})$ is 
\begin{equation}
{\cal E}(\alpha_i , \alpha_{j}) = \begin{dcases}  - (E_0 + \Delta) & \text{ if $|j - i| = N$} \\ - E_0 & \text{ if $|j-i| \neq N$}.  \end{dcases}
\label{eq:bin_eng}
\end{equation}
We term $E_0$ the "offset binding energy", and $\Delta$ the "energy advantage" of correct dimers. For simplicity, we will assume that the monomers and dimers are point particles with no rotational or vibrational properties. Also, apart from their binding, the monomers and the dimers are free particles which do not interact with one another. Therefore, the total energy of a microstate comes from the kinetic energies of the monomers and the kinetic energies and binding energies of the dimers. An example microstate for a non-gendered dimer system is shown in \reffig{non_gen_sys}. 

In order to study the thermal equilibrium properties of such a system, we need to construct its partition function. To build the partition function we must define the microstates of the system, the energy of a microstate, the various degeneracy factors relevant to defining a microstate, and how we will sum over all microstates. Given the definition of our system, a naïve choice for how to characterize the system microstate is to use a $2N\times 2N$ contact matrix ${\cal C}$ whose elements are defined according to
\begin{equation}
{\cal C}_{ij} = \begin{dcases} 1 & \text{ if dimer $(\alpha_i, \alpha_j)$ exists in system,} \\  0 & \text{ otherwise}. \end{dcases}
\label{eq:cij}
\end{equation}
With the elements ${\cal C}_{ij}$, we can then specify which monomers exist in isolation and which are bound together. From the constraints of the system, we can also infer that ${\cal C}_{ij}$ has no diagonal elements, is symmetric, and only has a single non-zero entry of $1$ in each column or row. Given \rfw{bin_eng} and \rfw{cij}, the energy of a particular microstate would then be 
\begin{align}
E\big( \{{\cal C}_{ij}\}\big) & = \sum_{i < j}^{2N} {\cal C}_{ij} {\cal E}(\alpha_i, \alpha_j) \mm
& = -E_0 \sum_{i < j}^{2N} {\cal C}_{ij} - \Delta  \sum_{i < j}^{2N} {\cal C}_{ij} \,\delta_{N, \,j-i}.
\label{eq:eng}
\end{align}
By the definition of the contact matrix in \rfw{cij}, the total number of dimers in the system is $\sum_{i<j} {\cal C}_{ij}$, and the total number of monomers is $2N- 2\sum_{i< j} {\cal C}_{ij}$. Presuming we are working under dilute-solution conditions, the monomers and dimers are non-interacting, and the phase-space degeneracy of a particular microstate ${\cal C}_{ij}$ can be accounted for by including factors of the ideal-gas partition functions for the appropriate number of monomers and dimers. If we have $N$ distinguishable and non-interacting point particles of mass $m_0$, the free-particle contribution to the partition function is 
\begin{equation}
Z_{\text{free}} = \left(\frac{V}{\lambda_0^3} \right)^N,
\label{eq:free}
\end{equation}
where $V$ is the volume of the system, and  $\lambda_0 =  {h}/{\sqrt{2\pi m_0 k_BT}}$ is the thermal de Broglie wavelength of a single monomer. There is no permutation correction in \rfw{free} because our particles are distinguishable. 
From \rfw{eng} and \rfw{free}, the partition function for the dimer system can be expressed as 
\begin{align}
Z_{N}(V, T,  E_0, \Delta) &= \sum_{\{{\cal C}_{ij}\}} \exp\left[-\beta  \sum_{i < j}^{2N} {\cal C}_{ij} {\cal E}(\alpha_i, \alpha_j) \right] \mm
&\,\,  \times \left(\frac{V}{\lambda_0^3} \right)^{2N- 2\sum {\cal C}_{ij}} \left(\frac{V}{(\lambda_0/\sqrt{2})^3}\right)^{\sum{\cal C}_{ij}}
\label{eq:naive_part}
\end{align}
\noindent where $\beta = 1/k_BT$,  the dimers have mass $2m_0$, $\sum {\cal C}_{ij}$ sums over indices $i<j$, and the microstate summation runs over all valid contact matrices for this system. 

Our larger objective is to derive an analytic form for the partition function and to then use this partition function to derive the thermal equilibrium conditions. But according to \rfw{naive_part}, in order to compute the partition function we have to enumerate and then sum over all valid contact matrices for this system. The set of possible contact matrices are all $2N\times 2N$ matrices that are symmetric, have no diagonal elements, and where each row's and each column's only non-zero element is 1. Finding a systematic way to enumerate such matrices is challenging enough, but further complicating the calculation is the way the binding energy \rfw{eng} changes contingent on which elements in ${\cal C}$ are non-zero. 

We can bypass these complications by expressing \rfw{naive_part} as a summation over states defined by the number of total dimers and number of correct dimers in the system. In terms of the contact matrix, we have 
\begin{equation}
k = \sum_{i <j}^{2N} {\cal C}_{ij}\,, \qquad m = \sum_{i < j}^{2N} {\cal C}_{ij}\, \delta_{N,\, j-i},
\end{equation}
as the number of total dimers and the number of correct dimers, respectively. Then, rather than defining and summing over all possible contact matrices, we need only sum over the possible values of $k$ and $m$ with the appropriate Boltzmann and degeneracy factors. In constructing the partition function, we define a state by a particular value of $k$ and $m$. \rfw{eng} indicates that the binding energy for such a state is $-kE_0 -  m \Delta $. Therefore, the partition function \rfw{naive_part} can be written as 
\begin{align}
Z_{N}(V, T, E_0, \Delta) & = \sum_{k=0}^N \sum_{m=0}^{k} \Omega_{N}(k, m)\,e^{\beta (kE_0 +m\Delta)}\mm
& \qquad \times \left(\frac{V}{\lambda_0^3} \right)^{2N-2k}  \left(\frac{V}{(\lambda_0/\sqrt{2})^3} \right)^k  \, 
\label{eq:sop_part}
\end{align}
where $\Omega_{N}(k, m)$ is the number of ways to construct a microstate with $k$ dimers, of which only $m$ are correct dimers. The task of computing the partition function now reduces to the task of computing the degeneracy factor $\Omega_{N}(k, m)$, and this calculation amounts to a problem of combinatorics. 

\subsection{Dance-Hall Problem \label{sec:dance}}
Determining $\Omega_{N}(k, m)$ generalizes beyond the constraints of this problem, and we can embed its definition in the answer to a less abstract problem. We phrase the problem as follows: 
\begin{quote}
$N$ pairs of people enter a dance hall. All people in the pairs separate, and people mingle with one another such that at some later time, some people are paired and other people are alone. At this later time, there are $k$ pairs of people on the dance floor, and of this set, there are $m$ pairs from the set of original pairs. How many ways can this happen? 
\end{quote}

The quantity $\Omega_{N}(k, m)$ is the answer to this question. To determine this quantity, we break it up into two factors: $\Omega_{N}(k, m)$ can be written as a product between the number of ways to select $m$ of the original pairs from the initial set of $N$ pairs and the number of ways to create, from the remaining $2(N-m)$ people, $k-m$ pairs which are not amongst the remaining $N-m$ original pairs. We thus have 
\begin{equation}
\Omega_{N}(k, m) = \binom{N}{m} a_{N-m, k-m}, 
\label{eq:omeg}
\end{equation}
where $a_{n, \ell}$ is the number of ways to form $\ell$ pairs from a set of $2n$ originally paired elements such that none of these $\ell$ pairs coincides with any of the original $n$ pairs. 

The quantity $\Omega_N(k, m)$ must satisfy a summation identity which we can use to check our final result. The total number of ways to form $k$ pairs out of a collection of $2N$ people (each of which can form a pair with any other person) is the number of ways to select $2k$ people to be amongst the pairs multiplied by $(2k-1)!! \equiv (2k)!/(2^k k!)$, the number of ways to rearrange the selected people amongst the pairs \cite{chuan1992principles}. Thus, upon summing \rfw{omeg} over all possible values of $m$ we should find
\begin{equation}
\binom{2N}{2k} (2k-1)!! = \sum_{m=0}^{k}\binom{N}{m} a_{N-m, k-m}.
\label{eq:omeg2}
\end{equation}

It is easy to calculate $a_{n, \ell}$ for a few representative values. For $\ell =1$, $a_{n, 1}$ is the number of ways to create a single pair that is not among the original $n$ pairs. In other words, $a_{n,1}$ is the difference between the number of ways to pair $2n$ objects and the number of original pairs:
\begin{equation}
a_{n, 1} = \frac{2n(2n-1)}{2} - n = 2n(n-1),
\label{eq:an1}
\end{equation}
For $\ell = n$, $a_{n, \ell}$ reduces to a solution to the "bridge couples problem" \cite{margolius2001avoiding}: The number of ways to completely rearrange $n$ paired people into $n$ new pairs such that none of these pairs is among the original collection is 
\begin{equation}
a_{n, n} = \sum_{j=0}^{n}(-1)^{j} \binom{n}{j}  (2n-2j-1)!!.
 \label{eq:bridge}
\end{equation}
For general $\ell$, we can find $a_{n, \ell}$ by applying the principle of inclusion and exclusion \cite{chuan1992principles}. We work through this derivation in \textit{SM Sec. C} and ultimately find 
\begin{equation}
a_{n, \ell} = \sum_{j=0}^{\ell}(-1)^{j} \binom{n}{j} \binom{2n-2j}{2\ell-2j} (2\ell-2j-1)!!. 
\label{eq:anl_fin}
\end{equation}
It is simple to check that \rfw{anl_fin} satisfies \rfw{bridge} and straightforward to check that it satisfies \rfw{an1}. 
To check \rfw{omeg2}, it is necessary to express \rfw{anl_fin} in terms of an integral as is done at the end of \textit{SM Sec. C}.

\subsection{Final Partition Function}

Expressing \rfw{omeg} in terms of the derived result \rfw{anl_fin}, we find that \rfw{sop_part} provides an exact partition function for our system of dimer-forming non-gendered monomers. However \rfw{sop_part} is not yet in its most reduced form because it is written as a summation over discrete indices. We can write this partition function in a form more responsive to the methods of calculus by using additional integration and combinatorial identities (see \textit{SM Sec. D} for details). In the end, we find the partition function 
\begin{widetext}
\begin{align}
Z_{N}(V, T, E_0, \Delta) = \frac{1}{2 \sqrt{\pi}\,\Gamma\left(N+1/2\right)}\left( \frac{V}{\lambda_0^{3}}\right)^{2N}\int^{\infty}_{0} \int^{\infty}_{0} dx\, dy\, \frac{e^{-x-y}}{\sqrt{xy}}\,\Big( {{\cal M}_{+}}^{2N} + {{\cal M}_{-}}^{2N}\Big),
\label{eq:final_part}
\end{align}
\end{widetext}
where 
\begin{equation}
{\cal M}_{\pm} \equiv \sqrt{x}\,\, \pm\,\, \left( \frac{2 \sqrt{2}\,\lambda_{0}^3 }{V}\right)^{1/2}e^{\beta E_0/2}\,\sqrt{y \,\Phi(x; \beta \Delta)}\, ,
\end{equation}
and 
\begin{equation}
\Phi(x; \beta \Delta) \equiv e^{\beta \Delta}+2x - 1,
\end{equation}
with $\Gamma$ being the Gamma function. \rfw{final_part} is an exact result and no mathematical approximations have been made in obtaining it. Thus it is valid for all $N$.

The advantage of expressing our original partition function \rfw{naive_part} as \rfw{final_part} is that, as an exponential integral, \rfw{final_part} is now amenable to approximation via Laplace's method, and we can use this method to obtain the equilibrium conditions of the system. First, given the appearance of $k$ and $m$ in \rfw{naive_part}, we can compute the average number of dimers with
\begin{equation}
\langle k \rangle = \frac{\partial}{\partial (\beta E_0)} \ln Z_{N},
\label{eq:kdef}
\end{equation}
and the average number of correct dimers with
\begin{equation}
\langle m \rangle = \frac{\partial}{\partial (\beta \Delta)} \ln Z_{N}.
\label{eq:mdef}
\end{equation}
We can use similar derivatives to compute the various elements of the covariance matrix for $k$ and $m$:
\begin{equation}
\left( \begin{array}{c c} \sigma_{k}^2 & \sigma_{km}^2 \\[0.5em] \sigma_{mk}^2 & \sigma_{m}^2 \end{array}\right) = \left( \begin{array}{c c} \partial^2_{\beta E_0} & \partial_{\beta E_0}\partial_{\beta \Delta} \\[0.5em]  \partial_{\beta \Delta}\partial_{\beta E_0} & \partial^2_{\beta \Delta}\end{array}\right) \ln Z_{N},
\label{eq:cov_mat}
\end{equation}
where $\sigma_{k}^2$ is the variance of the total number of dimers, $\sigma_{m}^2$ is the variance of the number of correct dimers, and $\sigma_{km}^2 = \sigma_{mk}^2$ is the covariance between the total number of dimers and the number of correct dimers. 

\rfw{kdef}, \rfw{mdef}, and \rfw{cov_mat} represent the main physical observables of this model, and computing these quantities will allow us to better characterize the various properties of the self-assembling dimer system. For example, we should be able to determine the conditions under which the energetic benefit for having a state of all correct dimers outweighs the entropic cost of not only having dimers rather than monomers but also of selecting the $N$ correct dimers out of a much larger set of incorrect dimers. Such conditions would constitute "regime" conditions for this system, and in order to find these conditions we first need to more specifically characterize the equilibrium properties of the system. 

\section{Equilibrium Conditions of Non-Gendered System \label{sec:eqbm}}
With the partition function \rfw{final_part}, we now have the main theoretical tool we need to explore the equilibrium properties of our system of non-gendered monomers. Our next task is to extract from this partition function physical information concerning the number of total dimers and the number of correct dimers. However, keeping \rfw{final_part} as an integral in the subsequent analysis would result in cumbersome integral expressions for both $\langle k \rangle$ and $\langle m \rangle$. It would be far simpler to approximate \rfw{final_part} as a function without an integral, and to then use this new function as a proxy for the partition function. 

Working towards this goal, we first rewrite \rfw{final_part} in a more suggestive form. Defining the effective free energy as 
\begin{align}
&\beta F_N(x, y;  V, T, E_0,  \Delta) \mm
&  = x + y + \frac{1}{2} \ln (xy)  - \ln \left({\cal M}_{+}^{2N} + {\cal M}^{2N}_{-} \right)+ \beta F_0(N,  V, T), 
\label{eq:fdef1}
\end{align}
where $\beta F_0(N, V, T)$ represents terms that are independent of the variables $x$ and $y$, we have  
\begin{align}
  &Z_{N}(V, T, E_0, \Delta)  \mm 
  & \quad = \int^{\infty}_{0} \int^{\infty}_{0} dx \,dy\, \exp\big[ - \beta F_N(x, y;  V, T, E_0,  \Delta) \big].
\end{align}
Next, by Laplace's method \cite{breitung2006asymptotic}, we can take $Z_N$ in the $N \gg 1$ limit to be dominated by the local maximum of its exponential integrand. We can then make the approximation  
\begin{align}
Z_{N}&(V, T,  E_0, \Delta) \simeq 2\pi  \left(\det H \right)^{-1/2} \exp\big[- \beta  F_N \big]\Big|_{x = \bar{x}, y = \bar{y} },
\label{eq:fapprox}
\end{align}
where $\bar{x}$ and $\bar{y}$ are the critical points of \rfw{fdef1} defined by 
\begin{equation}
\partial_{i} (\beta F_N)\Big|_{x = \bar{x}, y = \bar{y} } = 0, 
\label{eq:crit_conds}
\end{equation}
for $i = x, y$, and $H$ is the Hessian matrix with the elements
\begin{equation}
H_{ij} = \partial_{i} \partial_{j} (\beta F_N)\Big|_{x = \bar{x}, y = \bar{y} }.
\label{eq:hessian0}
\end{equation} 
In order for \rfw{fapprox} to be a valid approximation (and have an error of at most ${\cal O}(N^{-1})$), then $\bar{x}$ and $\bar{y}$ must not only satisfy \rfw{crit_conds}, but the Hessian matrix at these critical points must also   be positive definite \cite{hubbard2015vector}, namely, it must satisfy
\begin{equation}
\det H > 0\,, \quad \Tr H>0.
\label{eq:hessian}
\end{equation}
The two conditions \rfw{crit_conds} and \rfw{hessian} together ensure that $\beta F_N$ is at a local minimum at the critical points $\bar{x}$ and $\bar{y}$ and thus that it properly defines the thermodynamic equilibrium of the system. 

With the right side of \rfw{fapprox} we now have a closed form expression that we can use as a proxy for a our partition function. We can transcribe the mostly mathematical conditions defining $\beta F_N$ into physical results by using  \rfw{kdef}, \rfw{mdef}, and \rfw{fapprox}, to establish a system of equations between $\langle k \rangle$, $\langle m \rangle$, $\bar{x}$, and $\bar{y}$. In deriving these equations, we take \rfw{fdef1} evaluated at $x = \bar{x}$ and $y= \bar{y}$ to be the true free energy of this system \cite{Note1}. Solving this system, we obtain  equilibrium conditions written exclusively in terms of $\langle k \rangle$ and $\langle m \rangle$:
\begin{align}
\frac{4 \sqrt{2}\,\lambda_0^3}{V}\, e^{\beta E_0} &  =  \frac{\langle k \rangle - \langle m \rangle (1 - e^{-\beta \Delta})}{\big(N- \langle k \rangle\big)^2 } \label{eq:eqbm_k}\\[.75em]
\frac{e^{\beta \Delta}}{2} & = \langle m \rangle  \frac{N - \langle m \rangle (1- e^{-\beta \Delta})}{\langle k \rangle - \langle m \rangle (1- e^{-\beta \Delta})}.
\label{eq:eqbm_m}
\end{align}
In \textit{SM Sec. E.1}, we derive the conditions \rfw{eqbm_k} and \rfw{eqbm_m}, and in \textit{SM Sec. E.2} we ensure the validity of Laplace's method by checking that the relevant critical points satisfy \rfw{hessian}. To be precise, these equilibrium conditions have errors of the order of ${\cal O}\left( \langle k \rangle^{-1}\right)$ and ${\cal O}\left(N^{-1}\right)$, but we will take them to be exact in the subsequent analysis because these errors only become relevant when we are considering few particle systems or systems which are mostly composed of monomers. 

\rfw{eqbm_k} and \rfw{eqbm_m} tell us how the average number of dimers $\langle k \rangle$ and the average number of correct dimers $\langle m \rangle$ relate to each other and to system parameters like the number of particles, system volume, and the binding energies of correct and incorrect dimers. Their form is reminiscent of law of mass action equations---i.e., they have an energy dependent exponential term on one side and particle number ratios on the other---however, there are some important differences. For one, factors of  $(1-e^{-\beta \Delta})$ multiply the average number of correct dimers, and this is a feature which we will later see is important in deriving results for the $\Delta \to 0$ limit of the system. Moreover, in \rfw{eqbm_m} there is an $N$ dependent term which cannot be related to the typical particle number ratios of the law of mass action, but which we will see is important in defining the state of fully correct dimerization. 

With \rfw{cov_mat} we can calculate the covariance and variance relationships between the average number of dimers and the average number of correct dimers. Using the approximate free energy given in the \textit{SM Eq.(E7)} and evaluated at $x = \bar{x}$, $y = \bar{y}$, we find
\begin{align}
\sigma_{k}^2 & = \frac{1}{2N} \langle k \rangle \big(N - \langle k \rangle\big),  \label{eq:covs_1} \\ 
\sigma_{km}^2  & =  \frac{1}{2N} \langle m \rangle \big( N - \langle k \rangle \big), \\
\sigma_{m}^2 & = \langle m \rangle - \frac{\langle m \rangle^2}{2} \left( \frac{1}{\langle k \rangle} + \frac{1}{N} \right),
\label{eq:covs_2}
\end{align}
indicating, as one should expect, that the thermal fluctuations in our order parameters go to zero once the system becomes completely dimerized ($\langle k \rangle \simeq N$) and completely composed of all correct dimers ($\langle m \rangle \simeq N$). 

From here, we could attempt to solve the equilibrium conditions \rfw{eqbm_k} and \rfw{eqbm_m} and obtain explicit expressions for $\langle k \rangle$ and $\langle m \rangle$ as functions of temperature and other system parameters. However, as coupled quadratic equations, these conditions yield quartic equations for $\langle k \rangle$ and $\langle m \rangle$. There are methods for obtaining analytic solutions to quartic equations \cite{irving2013beyond}, but the general solutions are sufficiently complicated as to not be physically useful. So we instead solve these equilibrium conditions numerically. 

But before we pursue a numerical solution, we can still build understanding of the system by analytically considering two special cases: The case where correct dimers do not have a binding energy advantage over incorrect dimers, and the case where the offset binding energy is so large that all monomers have formed (not necessarily correct) dimers. 

\subsection{No Energy Advantage ($\Delta = 0$) \label{sec:no_adv} }

We consider the system without correct dimers having an energy advantage over incorrect dimers, namely the case where $\Delta =0$. For this case, we  define the system by the reaction equation 
\begin{equation}
\alpha_i + \alpha_j
\xrightleftharpoons[  ]{ } \alpha_i \alpha_j, \qquad \text{Binding Energy} = - E_0
\label{eq:rxnrate0a}
\end{equation}
where $- E_0$ is the binding energy of the forward reaction. The partition function for such a system can easily be written down by taking the appropriate limit of the partition function \rfw{final_part}. We find 
\begin{widetext}
\begin{equation}
Z_{N}(V, T, E_0, \Delta =0) =  \frac{1}{2\sqrt{\pi}}\left(\frac{V}{\lambda_0^3}\right)^{2N} \int^{\infty}_{0}dy\, \frac{e^{-y}}{\sqrt{y}} \left[\big(1+  \sqrt{2 \delta y }\,\big)^{2N} + \big(1- \sqrt{2 \delta y}\,\big)^{2N}\right],
\label{eq:E0zero}
\end{equation}
\end{widetext}
where 
\begin{equation}
\delta \equiv \frac{2 \sqrt{2} \, \lambda_0^3}{V} e^{\beta E_0}.
\end{equation}
To derive the equilibrium conditions for this system, we can apply Laplace's method to \rfw{E0zero} in a way similar to the method's application to \rfw{final_part}. However, doing so would lead to equilibrium conditions for $\langle k \rangle$ alone, since the parameter $\Delta$ (which defines $\langle m\rangle$ through $\langle m \rangle = \partial \ln Z_N /\partial(\beta \Delta)$) is absent. Alternatively, we can simply consider \rfw{eqbm_k} and \rfw{eqbm_m} for $\Delta =0$. Doing so, we find
\begin{equation}
\frac{4 \sqrt{2}\,\lambda_0^3}{V} \,e^{\beta E_0} =  \frac{\langle k \rangle}{\big(N- \langle k \rangle\big)^2 }\,, \qquad \langle m \rangle = \frac{\langle k \rangle}{2N}
\label{eq:eqbm_km2sp}.
\end{equation}
We note that it is the $(1-e^{-\beta \Delta})$ dependance of \rfw{eqbm_k} and \rfw{eqbm_m} that allows us to so easily take this $\Delta \to 0$ limit.

The equations in \rfw{eqbm_km2sp} equations have straightforward interpretations from the perspective of the law of mass action and basic counting. Identifying the concentration of monomers as $\text{[monomers]} = (2N - 2 \langle k \rangle)/V$ and the concentration of dimers as $\text{[dimers]} = \langle k \rangle /V$, we can write the first equation in \rfw{eqbm_km2sp} as 
\begin{equation}
{\sqrt{2} \,\lambda_0^3}\,e^{\beta E_0} = \frac{\text{[dimers]}}{\text{[monomers]}^2}, 
\label{eq:lma}
\end{equation}
which is reminiscent of a law of mass action interpretation of \rfw{rxnrate0a}. The left side of \rfw{lma} is off by a factor of $2$ from what we would precisely calculate using the law of mass action because a foundational assumption of our dimer system is that each $\alpha_i$ occurs once and is distinguishable from $\alpha_j$ for $j \neq i$, and such an assumption of distinguishability is not manifest in the simple "$\text{monomer} +\text{monomer} \xrightleftharpoons[  ]{ } \text{dimer}$" rendering of \rfw{rxnrate0a}. 

The second equation in \rfw{eqbm_km2sp} can be understood with a simple argument. If there is no energy difference between correct dimers and incorrect dimers, then the ratio between the average number of correct dimers and the average number of dimers should be equal to the ratio between the possible values of each. Given that there are $N$ possible correct dimers and $2N(2N-1)/2$ possible dimers, we should find that the ratio between the average number of correct dimers and the average number of dimers at thermal equilibrium is 
\begin{equation}
\frac{\langle m \rangle}{\langle k \rangle} = \frac{N}{2N(2N-1)2/2} = \frac{1}{2N-1}, 
\end{equation}
which, in the $N\gg1$ limit, is consistent with the second equation of \rfw{eqbm_km2sp}. 

\begin{figure*}[t]
\begin{centering}
\begin{subfigure}{0.3\textwidth}
\centering
\includegraphics[width=.8\linewidth]{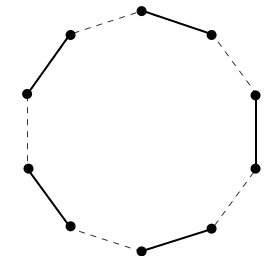}
	\caption{}
	\label{fig:m5}
\end{subfigure} 
\begin{subfigure}{0.3\textwidth}
\centering
  \includegraphics[width=.8\linewidth]{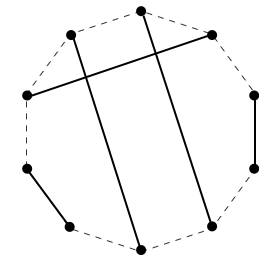}
	\caption{}
	\label{fig:m2}
\end{subfigure} 
\begin{subfigure}{0.3\textwidth}
  \centering
  \includegraphics[width=.8\linewidth]{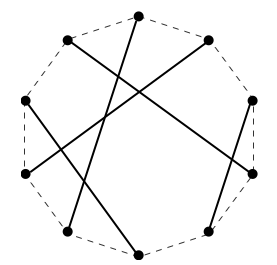}
	\caption{}
	\label{fig:m0}
\end{subfigure}
	\caption{Example microstates of a graph system with $2N=10$ vertices each of which has degree 1. The graph in (a) defines the lowest energy microstate with energy $E= - 5\Delta$. Each graph that has an edge not found in the lowest energy graph incurs an energy penalty $+\Delta$. Therefore, the graph in (b) has an energy $E= -2\Delta$, and the graph in (c) has an energy $E= 0$. Studying the equilibrium statistical physics of such a collection of graphs leads to the partition function in \rfw{znel} without the factor of $c^N$ and the additional corrections. \rfw{mval} indicates that the system assumes its lowest energy graph at or below the non-zero temperature $\Delta/\ln(10)$. }
	\label{fig:graph}
\end{centering}
\end{figure*}

\subsection{Complete Dimerization ($E_0 \gg k_BT$) \label{sec:all_dimer}}

If our dimer system had an offset binding energy that was much larger than the energy scale of thermal fluctuations, then the system would be entirely composed of dimers, and the corresponding thermodynamics would be determined by the combinatorics of correct and incorrect interactions. In such a situation, the only energy parameter relevant in defining the microstate of the system would be $\Delta$. In this $E_0 \gg k_BT$ limit, the partition function \rfw{final_part} reduces to 
\begin{align}
Z_N&(V, T, E_0 \gg k_BT, \Delta)\mm
& \quad = \frac{c^N}{\sqrt{\pi}}\int^{\infty}_{0} dx\, \frac{e^{-x}}{\sqrt{x}} \Big(e^{\beta \Delta} + 2x-1 \Big)^{N} + {\cal O}(c^{-1}),
\label{eq:znel}
\end{align}
where $c = (V/\lambda_0^3)e^{\beta E_0}$. Given that $\langle k \rangle = \partial \ln Z_N /\partial (\beta E_0)$, \rfw{znel} implies that $\langle k \rangle \simeq N$ as we expect for complete dimerization. Analyzing \rfw{eqbm_k} in this limit is difficult because of the divergence that occurs as $\langle k \rangle$ approaches $N$, but the second equation suffers no such divergence. Using $\langle k \rangle \simeq N$ in \rfw{eqbm_m} yields for $\langle m \rangle$
\begin{equation}
\langle m \rangle \simeq \frac{e^{\beta \Delta}}{2}. 
\label{eq:mval}
\end{equation}
At the highest temperatures, \rfw{mval} gives us the expected result that the system reduces to one of virtually no correct dimers, $\langle m \rangle \simeq 1/2$. However, given that $\langle m \rangle$ cannot exceed $N$, \rfw{mval} also implies that there is a finite temperature below which $\langle m \rangle \simeq N$, and hence at which all of the dimers in the system are correct. This temperature is $k_BT \simeq \Delta/\ln(2N)$. The fact that this temperature is non-zero for finite $N$ is important since such a result contradicts a potential expectation that complete order is only possible at zero temperature.  We do not call this behavior a phase transition since it disappears in the thermodynamic $N\to \infty$ limit, but it is clear that, like a phase transition, moving below this temperature results in behavior that cannot be fully captured by our analytic approximations. 

Finally, \rfw{znel} has a simple interpretation from the perspective of the statistical physics of graphs. We consider the set of graphs with $N$ edges and $2N$ vertices where each vertex has degree $1$. If we define one graph in this set as the lowest energy graph (with $E = - N\Delta$), and say that the system incurs an energy penalty $+\Delta$ whenever a graph has an edge not found in the lowest energy graph, then the partition function for the system is given by the first term in \rfw{znel} without the factor of $c^N$ and additional corrections. Moreover, \rfw{mval} indicates that below a temperature $\Delta/\ln(2N)$, the system settles into its lowest energy graph (\reffig{graph}).  In the next section, we will define the temperature at which fully correct dimerization occurs for arbitrary $\Delta$ and $E_0$, and we will see that $k_BT = \Delta/\ln(2N)$ is a special case of a more general result.

\section{Types and regimes of Dimer Systems \label{sec:phases}}

We say that our dimer system has undergone fully correct dimerization when the average number of correct dimers is equal to the average number of dimers, $\langle m \rangle = \langle k \rangle$. In this section, we use this definition to show that the dimer system can be categorized as one of two types. This categorization is based on analytic approximations for the temperature at which fully correct dimerization is achieved, and by plotting simulations and numerical solutions to \rfw{eqbm_k} and \rfw{eqbm_m} for these two system types, we find that the categorization also reflects a qualitative difference in the relationship between $\langle k \rangle$ and $\langle m \rangle$. With the intuition from these numerical analyses, we then define different physical regimes of the system (e.g., complete dimerization, partial dimerization, negligible dimerization etc.) and use the  $\beta E_0-\beta \Delta$ and $2N-V/\lambda_0^3$ parameter spaces to show that the two system types can access different regimes of self-assembly. 

\subsection{Type I and Type II Dimer Systems  \label{sec:1and2}}

\begin{figure*}[t]
\begin{centering}
\begin{subfigure}{0.32\textwidth}
\centering
\includegraphics[width=\linewidth]{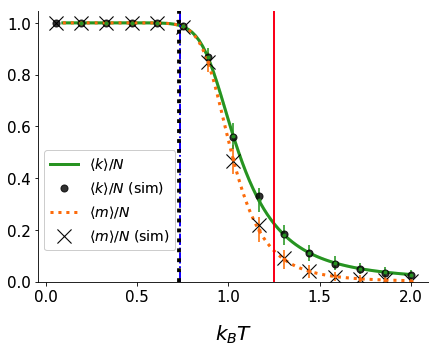}
	\caption{}
	\label{fig:typeIplot}
\end{subfigure} 
\begin{subfigure}{0.32\textwidth}
\centering
\includegraphics[width=\linewidth]{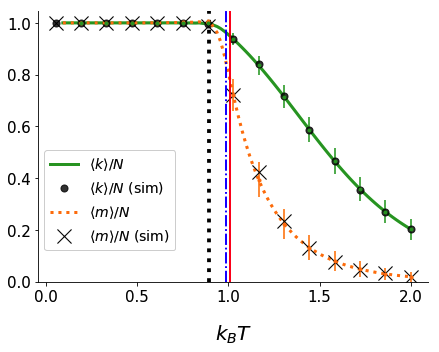}
	\caption{}
	\label{fig:typeIandIIplot}
\end{subfigure} 
\begin{subfigure}{0.32\textwidth}
  \centering
  \includegraphics[width=\linewidth]{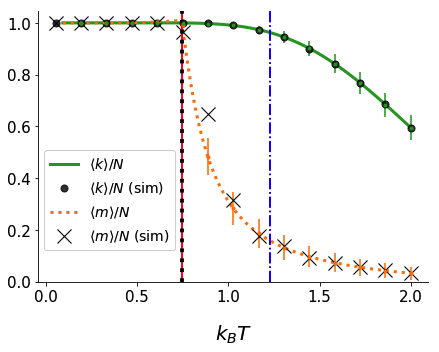}
	\caption{}
	\label{fig:typeIIplot}
\end{subfigure}
	\caption{(Color online) Numerical solutions to \rfw{eqbm_k} and \rfw{eqbm_m} and corresponding simulation results. We set $E_V = 10^{-3}$, $N = 50$, and defined all energies in units of $k_BT = 1.0$. The error bars are the standard deviations in $k$ and $m$ computed from \rfw{covs_1} and \rfw{covs_2}. In (a),  $E_0 = 4.15$ and $\Delta =  5.75$, and the system is Type I. In (b), $E_0 = 9.05$  and $\Delta = 4.65$, and the system is of indeterminate type. In (c), $E_0 = 14.00$ and $\Delta = 3.75$, and the system is Type II. The $\langle k \rangle$ and $\langle m\rangle$ numerical solutions are represented by solid green and dotted orange curves, respectively. The $\langle k \rangle$ and $\langle m \rangle$ simulation results are denoted by "$\bullet$" and "$\times$", respectively, and each point represents the average of 50 simulations where, for each simulation, the last 600 time steps of 30,000 were used to compute the ensemble average (see \textit{SM Sec. F} for details). Vertical lines correspond to $T_c$ (black dotted), $\TI$ (blue dashdotted), and $\TII$ (red solid). For Type I systems, $T_{c} \simeq \TI$, and for Type II systems, $T_c \simeq \TII$. In Type I systems, partially dimerized states can have mostly correct contacts, and in Type II systems partially dimerized states always have mostly incorrect contacts.}
	\label{fig:typeIandII}
\end{centering}
\end{figure*}

When our system is at high $T$ we can expect most of the monomers to exist alone such that $\langle m \rangle$, the average number of correct dimers,  and $\langle k \rangle$,  the average number of total dimers,  are both ${\cal O}(1)$. However, as we decrease the system temperature, we expect there to be a point at which $\langle m \rangle = \langle k \rangle$. At this point, we would say the system is in the regime of fully correct dimerization. At what temperature does the system enter this regime?

Imposing $\langle m \rangle = \langle k \rangle$ on both equations \rfw{eqbm_k} and \rfw{eqbm_m}, and presuming that this condition is first valid at the critical temperature $T_c$, we find that $T_c$ must satisfy 
\begin{equation}
\frac{\sqrt{2}\, \lambda_{0,c}^3}{V} \,e^{\beta_c(E_0 + \Delta)} \frac{\left(1- 2N e^{-\beta_c \Delta}\right)^2}{1-e^{-\beta_{c} \Delta}} = N-1/2
\label{eq:Tc}
\end{equation}
where $\lambda_{0,c} = h/\sqrt{2\pi m_0 k_BT_c}$ and $\beta_c = 1/k_BT_c$. Moreover, at this temperature, $\langle m \rangle$ and $\langle k \rangle$ assume the common value
\begin{equation}
\langle m \rangle = \langle k \rangle = \frac{N-1/2}{1-e^{-\beta_c \Delta}}.
\label{eq:mkeq}
\end{equation}
From \rfw{Tc}, we can show that $k_BT_c$ is bounded above by $\Delta/\ln (2N)$ which, together with \rfw{mkeq}, implies that, at $T= T_c$, $\langle k \rangle$ and $\langle m \rangle$ have a value between $N-1/2$ and $N$. Therefore, for this regime of fully correct dimerization, not only do all the dimers consist entirely of correct dimers, but all the monomers have formed dimers. 

For general parameter values, \rfw{Tc}  requires numerical methods to solve, but it is possible to find approximate analytical solutions in two limiting cases. In the case of large energy advantage for correct dimers ($\beta_c \Delta \gg 1$), the terms proportional to $e^{-\beta_c \Delta}$ go to zero, and we can solve for $T_c$ explicitly to find 
\begin{align}
k_BT_c & \simeq \frac{2}{3}(E_0 + \Delta)\left[W_{0} \left(\frac{E_0 + \Delta}{3 E_V}(2N)^{2/3}\right)\right]^{-1}\hspace{-0.2cm} +{\cal O}(N^{-1})\mm
& \qquad \equiv k_BT_{\text{I}},
\label{eq:T1}
\end{align}
where we defined 
\begin{equation}
E_V \equiv \frac{h^2}{2\pi m_0 V^{2/3}}, 
\end{equation}
as the effective energy of a free monomer of mass $m_0$ in a volume $V$, and where $W_0$ is the principal branch of the Lambert W function defined by the condition $W_0(xe^x) = x$ for $x>-1$ \cite{weisstein2002lambert}. Alternatively, in the case where the offset binding energy is large ($\beta_c E_0 \gg1$), the squared quantity in \rfw{Tc} must approach $0$ to compensate for its large coefficient, and we find 
\begin{equation}
k_BT_c \simeq \frac{\Delta}{\ln (2N)} \equiv k_BT_{\text{II}}.
\label{eq:T2}
\end{equation}
In practice, the solution to \rfw{Tc} cannot always be approximated by either $\TI$ or $\TII$, but in cases when it can, the corresponding thermal dependences for $\langle k \rangle$ and $\langle m \rangle$ are sufficiently different between these two limiting cases that it is appropriate to categorize these cases as two different system types. We define these two system types approximately as
\begin{equation}
\text{System Type} = \begin{dcases} \text{Type I} & \text{ for $T_{c} \simeq \TI $,}\\ \text{Type II} & \text{ for $T_c \simeq \TII $.} \end{dcases}
\label{eq:sys}
\end{equation}
For systems where $T_c$ cannot be approximated by either $\TI$ or $\TII$, we call the system type "indeterminate". 

In the following sub-sections, we explore this system categorization and the implications of \rfw{Tc} in two ways: First, using \rfw{sys} to categorize numerical solutions to \rfw{eqbm_k} and \rfw{eqbm_m}; second, constructing a parameter space plot of the solutions and using the system categorization to understand which spaces are accessible to Type I and Type II systems. 

\subsection{Numerical Solutions and Simulations  \label{sec:solsim}}

In \reffig{typeIandII}, we plot the numerical solutions to the equilibrium conditions \rfw{eqbm_k} and \rfw{eqbm_m} and compare these numerical results to simulation results for Type I, Type II, and indeterminate systems. The error bars in the plots are computed from \rfw{covs_1} and \rfw{covs_2}, and the system is simulated using a Metropolis-Hastings Monte Carlo algorithm with a set of moves chosen to ensure efficient exploration of the state space (see \textit{SM Sec. F} for details). 

The qualitative difference between Type I and Type II systems is apparent from comparing how $\langle k \rangle$ and $\langle m \rangle$ relate to one another for each system type. In both system types, when $T<T_c$, we find that $\langle k \rangle$ and $\langle m \rangle$ assume the value given by \rfw{mkeq}. But as we increase $T$ above $T_c$, Type I systems feature a soft transition from $\langle m \rangle \simeq N$ to $\langle m \rangle <N$ after which $\langle m \rangle$ closely shadows the behavior of $\langle k \rangle$, indicating that most of the dimers in such systems are correct. Conversely, Type II systems feature a sharp transition for $\langle m \rangle$ at $T \simeq \Delta/\ln (2N)$ followed by an exponential decline which drops $\langle m \rangle$ far away from the $\langle k \rangle$ value, indicating that most of the dimers in such systems are incorrect. The sharpness of the transition for Type II systems leads to relatively large fluctuations in $\langle m\rangle$ as shown by the larger discrepancy between simulation and analytic results in \reffig{typeIIplot} versus those in \reffig{typeIplot} and \reffig{typeIandIIplot}.

In general, above the critical temperature $T_c$, Type I systems have dimers that are dominated by correct contacts while Type II systems have dimers that are dominated by incorrect contacts.

\subsection{Parameter Space Plots \label{sec:param}}

\begin{figure*}[t]
\begin{centering}
\begin{subfigure}{0.49\textwidth}
\centering
\includegraphics[width=\linewidth]{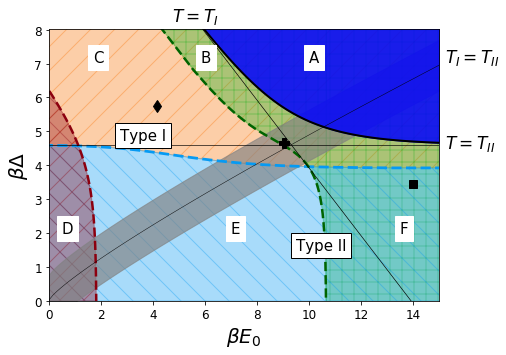}
	\caption{}
	\label{fig:E0delplot}
\end{subfigure} 
\begin{subfigure}{0.49\textwidth}
\centering
\includegraphics[width=\linewidth]{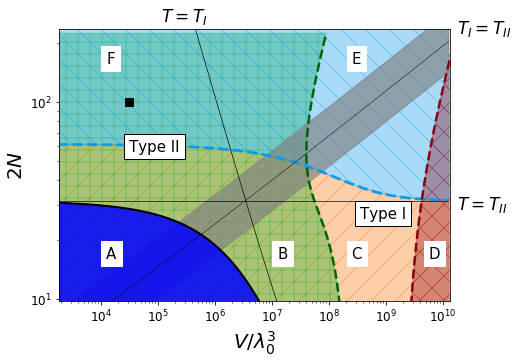}
	\caption{}
	\label{fig:2NVplot}
\end{subfigure} 
	\caption{(Color online) Parameter space regimes of dimer system. In (a), we set $E_V = 10^{-3}$ and $N = 50$, and in (b), we set $E_0 = 14.00$ and $\Delta = 3.45$; in both (a) and (b), we set $k_BT =1.0$. Each region is defined by solutions to \rfw{eqbm_k} and \rfw{eqbm_m} satisfying the following: (A)  fully correct dimerization (\rfw{mkeq}); (B) nearly complete dimerization with mostly correct contacts; (C)  partial dimerization with mostly correct contacts; (D)  negligible dimerization; (E) partial dimerization with mostly incorrect contacts; (F) nearly complete dimerization with mostly incorrect contacts. The curve bounding region A in (a) and (b) is, respectively, the function $\beta \Delta(\beta E_0)$ and the function $2N(V/\lambda_0^3)$ found from analytic solutions to \rfw{Tc}. The solid lines are functions computed from their respectively labeled conditions. The grey diagonal strip in (a) and (b) defines a region in which the system type is indeterminate; above or below the strip, the system is more clearly of Type I or Type II. The markers $\blacklozenge$, \tbf{\Large{+}}, and $\rule[0pt]{6pt}{6pt}$  correspond, respectively, to (a), (b), and (c) in \reffig{typeIandII} at $k_BT= 1.0$. Only Type I systems can be partially dimerized and mostly correct while only Type II systems can be nearly completely dimerized and mostly incorrect. }
	\label{fig:phase_dimer}
\end{centering}
\end{figure*}

In \rfw{mkeq}, we took the relationship $\langle m \rangle = \langle k \rangle$ to define the fully correct dimerization regime of the dimer system. This regime is evident in all the plots in \reffig{typeIandII} for $T\leq T_c$, but these plots also show that there are many different relationships between $\langle k \rangle$ and $\langle m \rangle$ that we can use to define various regimes of dimer assembly. It is easiest to get a sense of these regimes with parameter space plots.

\reffig{E0delplot} and \reffig{2NVplot} depict, respectively,  $\beta E_0-\beta \Delta$ and $2N-V/\lambda_0^3$ parameter spaces for the dimer system with $N$ and $V$ fixed in the former and $E_0$ and $\Delta$ fixed in the latter. A system at a particular temperature and with particular parameter values is located at a specific point on either parameter space plot. For example, the $k_BT=1.0$ values of $\langle k \rangle$ and $\langle m \rangle$ in the plots \reffig{typeIplot}, \reffig{typeIandIIplot}, and \reffig{typeIIplot} are represented, respectively, as $\blacklozenge$, \tbf{\Large{+}}, and $\rule[0pt]{6pt}{6pt}\,$ markers in \reffig{E0delplot}, and the $k_BT=1.0$ values of $\langle k \rangle$ and $\langle m \rangle$ in \reffig{typeIIplot} are represented by  $\rule[0pt]{6pt}{6pt}\,$ in \reffig{2NVplot}. We emphasize that because our results are derived in the $N\gg1$ limit, the properties outlined for \reffig{2NVplot} become less accurate descriptions of the original system for lower values of $2N$.

The solid straight lines are the parameter space expressions of the conditions $T=\TI$, $T = \TII$, and $\TI = \TII$ given the definitions in \rfw{T1} and \rfw{T2}. If we take a system at a certain temperature to be defined by a point in \reffig{E0delplot} or \reffig{2NVplot}, then decreasing the system temperature brings the point closer to region $A$. Because the region boundaries are themselves temperature dependent, the sizes and extents of the regions also change as we change the system temperature. See Fig. (S1) in \textit{SM Sec. G} for a depiction of how the plots in Fig. 5 change as we decrease the value of $k_BT$. We define a system as Type I or Type II according to whether decreasing the system temperature leads the point representing the system to enter region A (fully correct dimerization region) at a point at which either the $\TI$ or $\TII$ line can approximate the region A boundary. The temperatures $\TI$ and $\TII$ must be sufficiently distinct for this categorization to be non-ambiguous and so the grey regions in both plots of \reffig{phase_dimer} define approximate regions where $\TI \simeq \TII$ and hence where the system type is indeterminate.

In the parameter space plots, we define six regimes that an arbitrary dimer system can be in at a given temperature. 
\begin{enumerate} [(A)]
\item \textit{Fully correct dimerization: }All monomers exist in dimers and all dimers are correct; \rfw{mkeq}, $1-1/2N<\langle k \rangle/N = \langle m \rangle/N < 1$. 
\item \textit{Nearly complete dimerization with mostly correct contacts:} Almost all the monomers exist in dimers, and most of these dimers are correct; $\langle k \rangle/N > 0.95$; $\langle m \rangle/\langle k \rangle >0.5$; $\langle k \rangle \neq \langle m \rangle$.
\item \textit{Partial dimerization with mostly correct contacts:} Monomers have only partially dimerized, and most of these dimers are correct; $0.05< \langle k \rangle/N < 0.95$; $\langle m \rangle/\langle k \rangle >0.5$.
\item \textit{Negligible dimerization:} Few of the monomers exist in dimers; $\langle k \rangle/N < 0.05$.
\item \textit{Partial dimerization with mostly incorrect contacts:} Monomers have only partially dimerized, and most of these dimers are incorrect; $0.05< \langle k \rangle/N < 0.95$; $\langle m \rangle/\langle k\rangle <0.5$.
\item \textit{Nearly complete dimerization with mostly incorrect contacts:} Almost all the monomers exist in dimers, and most of these dimers are incorrect. $\langle k \rangle/N > 0.95$; $\langle m \rangle/\langle k \rangle <0.5$.
\end{enumerate}

The dotted line boundaries in \reffig{E0delplot} and \reffig{2NVplot} are defined by somewhat arbitrary limiting values for $\langle k \rangle$ and $\langle m \rangle$ (e.g., $\langle k \rangle/N <0.10$ and $\langle k \rangle/N > 0.90$ could respectively have been used to define negligible and nearly complete dimerization), and thus transitioning across such boundaries occurs smoothly as "crossover", rather than as "phase", transitions. However, the boundary surrounding region A is unambiguously defined by \rfw{Tc}, and transitioning across this boundary by decreasing $T$  below $T_c$ fixes $\langle m \rangle$ and $\langle k \rangle$ at the value given in \rfw{mkeq}. For Type I systems, this $T=T_c$ transition occurs smoothly (\reffig{typeIplot}), but for Type II systems the transition occurs sharply (\reffig{typeIIplot}) corresponding to an apparent discontinuity in $\partial \langle m \rangle/\partial T$ and thus suggesting the appearance of a phase transition. However, this transition occurs at an $N$ dependent temperature that goes to zero in the thermodynamic limit, and thus does not fulfill the standard definition of a phase transition. 

Echoing an assertion made in the previous section, \reffig{E0delplot} and \reffig{2NVplot} show that Type I and Type II systems exhibit regimes of behavior exclusive to each type. When monomers are partially dimerized in a Type I system, most of the dimers can consist of correct contacts, while when monomers are partially dimerized in a Type II system, most of these dimers always consist of incorrect contacts. 

These parameter space plots allow us to immediately see a few properties of the dimer system not evident in the solution plots. First, from the regime definitions and the line representing the $T= \TII$ condition in both  \reffig{E0delplot} and \reffig{2NVplot}, we see that $\beta \Delta > \ln (2N)$ (or, equivalently, $2N < e^{\beta \Delta}$) appears to be a sufficient but not necessary condition for an arbitrary system's dimers to be mostly composed of correct dimers. Therefore, the dimers in a system are mostly correct if the number of distinct monomers in the system is less than $e^{\beta\Delta}$.

Second, in \reffig{E0delplot} we see the expected result that the system only enters the fully correct dimerization regime when $\Delta \gg k_BT$ and $E_0 \gg k_BT$. This makes qualitative sense because a value of $E_0$ much larger than the energy scale of thermal fluctuations is needed for dimers to be able to form, and, similarly, a large value of $\Delta$ ensures that correct dimers are privileged over incorrect dimers. 

However, in \reffig{2NVplot} we have a possibly unexpected result: It is only the lower left corner of the $2N-V/\lambda_0^3$ parameter space that contains the fully correct dimerization regime. This suggests that it is the absolute values of both particle number and volume, rather than just their ratio encoded in number density, that determine whether fully correct dimerization is possible. This result might be unexpected since reaction equations similar to those defining our dimer system (i.e., similar to \rfw{rxnrate}) are often studied by considering reactant number densities in the form of concentrations. Experience with such analyses leads one to expect that limits on number density are the only relevant criteria for constraining whether correct dimerization is achieved. But now we see that a statistical mechanics analysis suggests otherwise. We interpret this result in the next section. 

\section{Inequalities for Assembly and Type \label{sec:ineq}}

Having constructed the parameter spaces in \reffig{phase_dimer}, we now pursue two goals: A qualitative interpretation of the analytical conditions constraining the fully correct dimerization regime, and a more precise way to define the separation between Type I and Type II systems. We pursue the first goal by finding necessary but not sufficient conditions for a system to be in the fully correct dimerization region of parameter space and then by using these conditions to motivate the more conceptual labels of "search-limited" and "combinatorics-limited" for Type I and Type II systems, respectively. We pursue the second goal by deriving and interpreting necessary but not sufficient conditions for a system to be of Type I. 

\subsection{Limits of Fully Correct Dimerization}

In \reffig{E0delplot} and \reffig{2NVplot},  region A defines the parameter space for which a dimer system is in the regime of fully correct dimerization. A necessary and sufficient condition for the system to be in this regime is $T< T_c$ where $T_c$ is given by the solution to \rfw{Tc}. The complexity of \rfw{Tc} makes this condition difficult to interpret physically, but the solid lines in the parameter space plots, corresponding to $T= \TI$ and $T= \TII$, allow us to state two necessary but not sufficient conditions that have clearer physical interpretations. 

From \rfw{Tc},  \rfw{T1}, and \rfw{T2}, we can show $T_c < \TI, \TII$. Thus, a necessary condition for the achievement of the fully correct dimerization regime is that $T< \TI$ and $T < \TII$. Using \rfw{T1} and \rfw{T2} to translate the $T< \TI$ and $T< \TII$ inequalities into physical limits on volume and particle number, we find that they correspond, respectively, to 
\begin{equation}
NV <  \sqrt{2}\,\lambda_0^3\,e^{\beta(E_0 + \Delta)},
\label{eq:search}
\end{equation}
and 
\begin{equation}
2N < e^{\beta \Delta}. 
\label{eq:comb}
\end{equation}
where, consistent with the $N\gg1$ limit, we dropped the ${\cal O}(N^{-1})$ term in \rfw{T1}. In \reffig{E0delplot}, \rfw{search} and \rfw{comb} are satisfied when a system exists to the right of the $T=\TI$ line and above the $T = \TII$ line.  In \reffig{2NVplot}, \rfw{search} and \rfw{comb} are satisfied when a system exists to the left of the $T = \TI$ line and below the $T = \TII$ line. 
Since the fully correct dimerization region exists within these limits in both figures, \rfw{search} and \rfw{comb} are necessary but not sufficient conditions for fully correct dimerization. Also, although they both contain the parameter $N$, \rfw{search} and \rfw{comb} are independent of one another. 

A system only satisfies \rfw{search} if it has binding energies $E_0$ and $\Delta$ which are strong enough for all $2N$ monomers to find and bind to one another in the volume $V$. We thus term \rfw{search} a "search-limiting" condition for the dimer system. A system only satisfies \rfw{comb} if it has an energy advantage $\Delta$ which is strong enough that the completely correct configuration of dimers is thermodynamically preferred over all the other combinatorially more numerous incorrect configurations. We thus term \rfw{comb} a "combinatorics-limiting" condition.

We can think of Type I systems as being "search-limited" since in such systems $\Delta$ is sufficiently large that correct dimers can overcome their combinatorial disadvantage, and, therefore, the primary limiting factor in creating correct dimers is the ability of the relevant monomers to find one another, i.e., satisfying \rfw{search}. Similarly, we can think of Type II systems as being "combinatorics-limited" since in such systems $E_0$ is sufficiently large that monomers can find one another, and the primary limiting factor in creating correct dimers is the need to overcome their combinatorial disadvantage, i.e., satisfying \rfw{comb}. 

It may seem strange that the inequality \rfw{search} is said to define the search-limits of dimer assembly and yet it makes no reference to the number density of the system. Shouldn't high number density be a requirement for monomers to be able to find one another in their volume? The answer depends on the properties of the monomers comprising the system. Number density is mainly relevant if the dimers formed from associating monomers are all identical, and the monomers exist in multiple copies which are uniformly distributed in the constituent volume. In such cases, dimerization occurs if the monomers can find one another, and since the reactants are uniformly distributed throughout their volume, the only factor constraining whether they are able to find one another is how many of these monomers are in a particular region of their larger space. Thus, only density is relevant.  

But for our dimer model, each of the $2N$ monomers exists as a single-copy, and all of the dimers are distinct. In order for the system to assume the fully correct dimerization regime, each monomer must ignore the $2N-2$ other monomers that are not its optimal binding partner and find the optimal partner in the volume $V$. Increasing the number of distinct monomers makes a successful search less likely since there are more spurious potential binding partners, as does increasing the system volume since there is a larger space to search within. Therefore, both $N$ and $V$ should have upper limit constraints. However, why is it their product $NV$ that has an upper limit constraint given in \rfw{search}? One answer is that particle number and volume are not independently constrained for a successful search. For example, a large volume and a small number of particles is just as harmful to a successful search as is a small volume and a large number of particles; in both cases a monomer still has to wade through a large number of various states---defined by possible position states or potential monomer binding partners---before it finds its optimal partner. Therefore the search limits on particle number become more stringent as the volume increases as do the search limits on volume when the particle number increases. Thus, it is their product which is constrained.

\subsection{Limits of System Type}

According to \rfw{sys} we categorized a dimer system as Type I or Type II contingent on how close $T_c$ was to either $\TI$ or $\TII$. This definition was necessarily approximate since the distinction between these two system types is a qualitative one which smoothly disappears as our system moves closer to the $\TI=\TII$ lines in \reffig{E0delplot} and \reffig{2NVplot}. But because of how $\TI$ and $\TII$ relate to one another in the two system types, we can rephrase the definition without explicit reference to how either relates to $T_c$. 

When $\TI$ and $\TII$ are not approximately equal, the critical temperature $T_c$ ends up being well approximated by the lower of the two values as is seen in \reffig{typeIplot} and \reffig{typeIIplot}. For Type I systems, the lower value is always $\TI$ and for Type II systems the lower value is $\TII$. Therefore, another way to define the system types is as
\begin{equation}
\text{System Type} = \begin{dcases} \text{Type I} & \text{ for $\TI < \TII $,}\\ \text{Type II} & \text{ for $\TI > \TII$.} \end{dcases}
\label{eq:sys_alt}
\end{equation}
where this definition is only unambiguous if $\TI$ and $\TII$ are not approximately equal. It is this phrase "not approximately equal" that makes this alternative definition (like the original definition \rfw{sys}) a qualitative one. However, this definition can be used as a guide to write a necessary but not sufficient condition for whether a system is of a particular type. 

\rfw{sys_alt} states that in order for a system to be of Type I, we must have $\TI < \TII$. This inequality alone is a necessary but not sufficient condition for the system to be of Type I. For example, \reffig{typeIandII} satisfies $\TI< \TII$, but its system type is ambiguous. Still, we can consider how this condition constrains the parameter space for this system. We rewrite this inequality in terms of a maximum number of distinct monomers for a Type I system. Using \rfw{T1} and \rfw{T2} in $\TI <  \TII$ and noting that, by the monotonicity property of the Lambert W function, if $W_0(X)>k$, then $X> k e^{k}$, we can show that $\TI< \TII$ implies 
\begin{equation}
2N < \exp\left[\frac{3\Delta}{2E_0}W_{0} \left(\frac{E_0}{3E_V}\right)\right].
\label{eq:2N_ineq}
\end{equation}
\rfw{2N_ineq} corresponds to the region in \reffig{2NVplot} that is below the $\TI=\TII$ line. Thus, if a dimer system can be categorized as Type I, then the number of distinct monomers it contains must satisfy \rfw{2N_ineq}.

\rfw{2N_ineq} is equivalent to a bare statement of the $\TI < \TII$ condition. However, unlike the $\TI<\TII$ condition, it presents constraints on $2N$ in terms of a closed-form expression and is thus easier to interpret. Taking $\Delta \ll E_0$ in \rfw{2N_ineq}, leads to a lower limit on the number of particles in the system. This makes sense because a smaller energy advantage for correct contacts means the system must have a smaller number of distinct monomers in order to avoid the prevalence of incorrect contacts which would push the system to be Type II. For larger volumes $V$, $E_V$ becomes smaller and \rfw{2N_ineq} indicates that the maximum value of $2N$ increases. This result is consistent with the fact that, by \rfw{T2}, increasing $2N$ decreases $\TII$: Since it is the positive difference between $\TII$ and $\TI$ that leads a system to be characterized as Type I, a decrease in $\TI$ through an increase in $V$ can be paired with a decrease in $\TII$ through an increase in $N$, with the system still maintaining its Type I status. It is true that increasing $N$ also decreases $\TI$, but because $W_0(x)$ varies more slowly than $\ln (x)$ this decrease occurs more slowly than the corresponding decrease in $\TII$. 

\rfw{2N_ineq} is a conceptually and analytically simple criterion for determining whether a dimer-system can be categorized  as Type I. Satisfying \rfw{2N_ineq} does not guarantee that the system is Type I, but failing to satisfy it guarantees that the system is not Type I. In the next section, we will use this criteria to determine whether various biomolecular systems have biophysical properties consistent with those of Type I dimer systems. 

\section{Biomolecular Systems \label{sec:bio}}

In this section we consider three systems whose properties approximately match the assumptions underlying the non-gendered or the gendered dimer models (the latter of which is outlined in the Appendix): The assembly of ssDNA into dsDNA, the specific and non-specific interactions between transcription factors and DNA, and the dimerization of distinguishable monomeric proteins into dimers (\reffig{dimer_int}).

There are some important differences between the model's assumptions and the properties of these real systems. 

First, we assumed that each monomer species exists in a single copy in the system. This assumption does not mirror the properties of real biomolecular systems which often have multiple copies, with different copy numbers, for important biomolecules. We take our model to approximate the behavior of systems with many different monomers but where the copy-numbers of each monomer are sufficiently similar and are uniformly distributed that we can consider a small region of the system to have a single-copy of each monomer type. In \refsec{discuss2} we will state a formulation of the non-gendered problem which better takes into account differences in particle number, and we will mention issues relevant to the solution.

Second, in developing the dimer model, we have employed the dilute-solution approximation throughout in which the monomers and dimers are presumed to be point-like and non-interacting. But, in real biomolecular systems, one would expect volume exclusion and intermolecular interactions to lead to deviations from ideal behavior. In \refsec{discuss2} we will comment on how we can make up for this limitation by extending the model, but for the current analysis we just acknowledge that the model only approximates the interaction properties of the monomers and dimers in the proposed real systems.

Third, our model uses only two parameters to define the binding energy matrix of $2N$ distinct monomers, whereas actual systems of distinct interacting proteins or strands of DNA would have more complicated binding interactions even if such interactions could be cleanly divided into correct and incorrect bindings. Consequently, in order to frame the properties of biomolecular systems in terms of model parameters, we use average energy scales representative of the systems of interest as approximations for $E_0$ and $\Delta$. 

Finally, in real biomolecular dimer systems, there are often rotational and vibrational contributions to entropy \cite{finkelstein1989price} which, in a more complete theoretical treatment, would have been accounted for in our dimer partition function \rfw{sop_part}. Because our model only takes into account the translational entropy of the dimers, when given biophysical data on binding free energies, we will take $E_0$ and $E_0+\Delta$ to be approximated by the provided binding free energies minus an estimated translational entropy contribution to those free energies. In this sense, the binding energy parameters of our model are "effective" binding energies obtained by averaging over the various unaccounted for internal microstates of the dimer, but are not directly associated with a measurable quantity. Carefully incorporating rotational and vibrational contributions into the partition function \rfw{sop_part} would lead to equilibrium conditions with different temperature dependences than those in \rfw{eqbm_k} and \rfw{eqbm_m}, and thus different conditions for Type I and Type II dimer systems. Thus taking $E_0$ and $E_0+\Delta$ to approximate these unaccounted for entropies amounts to an additional approximation in which we are ignoring the temperature dependence of these entropies. All binding energy calculations are found in the \textit{Supplementary Code}. 

In the subsequent sections, we will have two main goals: First, to use \rfw{search}, \rfw{comb}, and estimates of biophysical parameters for various biomolecular systems to determine how the diversity of monomers in the system would need to be constrained in order for fully correct dimerization to be accessible at physiological temperatures. Second, to determine whether the system is a Type I (search-limited) or Type II (combinatorics-limited) dimer system, and thus whether partially dimerized systems are dominated by correct contacts in these systems. Completing the first goal provides us with the information for the second goal: According to \reffig{2NVplot}, if a system satisfies \rfw{comb} but not \rfw{search}, then the system is of Type I, but if a system satisfies \rfw{search} but not \rfw{comb} then the system is of Type II. We will also use \rfw{2N_ineq} to affirm these system categorizations. 

\subsection{ssDNA-ssDNA interactions}

Within a cell, dsDNA never spontaneously separates into ssDNA, but in polymerase chain reactions (PCR), solutions containing copies of a single dsDNA sequences are heated to high enough temperatures that the strands can separate. In a prepared system consider having (instead of multiple copies of a single sequence of dsDNA as in PCR) $N$ different sequences of dsDNA which, when heated to high enough temperatures, separate into $N$ ssDNA segments and $N$ associated complementary segments (\reffig{ssdna}). 

This system is contrived from a biological perspective but provides a simple playground in which to study the predictions of the dimer model. What insights do the physical properties of the non-gendered dimer model provide for such a system of ssDNA and dsDNA? One relevant question is whether such a system is a Type I or a Type II dimer system. 

Take a single ssDNA segment to have 20-nucleotide bases, a length which is within the range of standard lengths of primers in a typical PCR  \cite{innis1999optimization}.  In the language of the model, each $\alpha_k$ for $k = 1, \ldots, N$, represents one ssDNA fragment and $\alpha_{N+k}$ represents the corresponding complementary fragment. Because each $\alpha_k$ is presumed distinct, we require that none of the ssDNA is self-complementary, and hence each is different from its complementary strand. We will assume binding occurs in an all-or-nothing fashion and that the bubbles that exist in real strands \cite{fei2013watching} are not present. The reaction equation for this system is 
\begin{equation}
\text{ssDNA}_{k} + \text{ssDNA}_{\text{comp}, k} \xrightleftharpoons[ ]{}  \text{dsDNA}_{k}
\end{equation}
where $k=1, \ldots, N$. 

Since only complementary ssDNA fragments can form dsDNA, there is no binding energy favorability between non-complementary ssDNAs, and so we can take $E_0=0$. From this condition alone, \reffig{E0delplot} suggests that such a system of interacting ssDNA is trivially of Type I, since a non-zero value of $\Delta$ and a zero value of $E_0$ would place the system well above the $\TI = \TII$ line. 

Still, we can consider what estimates for binding energies imply about the number of distinct ssDNA that can exist in such a system.  A representative binding free energy between complementary strands was found as follows: $10^6$ 20-base sequences of ssDNA (where the bases A, G, T, and C were equally probable) were randomly generated, the binding free energy for each with its corresponding complement was computed, and the result was averaged over all sequences . An experimentally calibrated and cross-referenced formula given in \cite{santalucia1998unified} was used to compute these free energies, assuming a 50 mM Na$^{+}$ surrounding solution (see \textit{Supplementary Code} for implementation details). The average free energy yielded an estimate for the binding energy parameter: $\Delta \simeq 31.5$ kcal/mol. From the fact that a nucleotide base pair has a mass of about 650 daltons,  the mass of a 20-base ssDNA was taken to be $m_0 = 6.5\text{ kDa}$ \cite{metzler2001biochemistry}. The system temperature was taken to be the physiological temperature $T = 310.15$ K. 

With these parameters \rfw{search} and \rfw{comb} yield, respectively, 
\begin{equation}
NV < 4.2 \times 10^{4} \text{ $\mu$m}^3, \qquad 2N <  1.6 \times 10^{22},
\end{equation} 
Since a 20-base pair ssDNA can have at most $4^{20} \approx 10^{12}$ distinct sequences, the combinatorial condition on $N$ is automatically satisfied, and it is thus the search condition $(NV)_{\text{max}}$ which limits the achievement of fully correct dimerization in this conjectured system. Moreover, taking $E_0 \to 0$ in the necessary condition \rfw{2N_ineq}, yields $2N < \exp(\Delta/2E_0) \approx \exp(10^{13})$ which is practically infinite and more than satisfied for the possible values of $N$ in the system. Therefore, this system is indeed of Type I  and is a search-limited dimer system.

\subsection{Transcription factor-DNA Interactions}

Transcription factors (TFs) are proteins that bind to DNA and regulate a gene's transcription into mRNA and thus how much protein is produced from that gene \cite{alberts2013essential}. Given their importance in gene regulation networks and the specificity of their functions, TFs must attach to precise regions of DNA which they select out of a combinatorial sea of other binding regions (\reffig{protein_dna}). A TF finding its intended DNA target is said to bind to it "specifically" while bindings to all other targets are considered "non-specific" \cite{jen1997protein}. 

Let's say we have $N$ different TFs in a system together with their corresponding $N$ DNA binding sites. The association and dissociation reaction for this system can be written as 
\begin{equation}
\text{TF}_{k} + \text{DNA}_{\ell} \xrightleftharpoons[ ]{}  \text{(TF-DNA)}_{k\ell}
\end{equation}
where $k, \ell =1, \ldots, N$. We want to use the biophysical parameters defining TF-DNA systems to consider what our model states about the diversity constraints of these systems. 

First, a system of interacting TFs and DNA sites is gendered because there are two types of interacting units and because we take the interactions to occur between respective members of the two types rather than within the same type (See \reffig{gen_sys} for an example of a gendered dimer system microstate). Also, since the DNA strand is fixed relative to the TFs, the system is more like a system of free monomers interacting with fixed binding sites rather than a system of dimer-forming monomers. Consequently, the reduced mass $\mu$ of the dimers becomes the mass of the motile monomer (i.e., the mass of the TF), and the qualitative picture we associate with the system is more akin to \reffig{fixed_gen_sys} in the Appendix than to \reffig{gen_sys}.

In \cite{jen1997protein}, Jacobsen lists 12 proteins (including endonucleases, repressors, and activators) with their respective protein-DNA association constants for specific and non-specific contacts under various conditions. Converting these association constants to binding free energies, and subtracting translational entropies to estimate our binding energy parameters $E_0$ and $\Delta$, we find $E_0 \simeq 22.9$ kcal/mol and $\Delta \simeq 6.4$ kcal/mol. We take the mass of a transcription factor monomer to be $m_{\text{TF}} \simeq 64 \text{ kDa}$, a typical protein mass \cite{milo2015cell}, and we take $T = 310.15$ K. 

From these parameter values, we find that gendered analogs of \rfw{search} and \rfw{comb} (given in \rfw{search_1} and \rfw{comb_1}, respectively) yield
\begin{equation}
NV< 2.7\text{ $\mu$m}^3, \qquad N< 3.2 \times 10^4.
\label{eq:limits_TF}
\end{equation} 
Both of these results establish limits on the maximum diversity of TFs needed for fully correct dimerization to be achievable at physiological temperatures, but the condition that establishes more stringent limits for a particular volume is what ultimately defines whether the system is of Type I or Type II. The authors of \cite{perez2000repertoire} estimate that there are about $N=3 \times10^2$ different TFs in \textit{E. coli}, a value which, for the \textit{E. coli} volume $1\text{ $\mu$m}^3$, satisfies the $(N_{\text{max}})$ condition but not the $(NV)_{\text{max}}$ condition. Thus, the $(NV)_{\text{max}}$ condition, derived from $T< \TI$, establishes the stronger limit on TF diversity for a $1\text{ $\mu$m}^3$ volume system, and we can conclude that this system is a Type I (i.e., search-limited) dimer system. Moreover, given our parameter values, we find that the gendered analog of \rfw{2N_ineq} (given in \rfw{N_ineq}) yields $N \lesssim 10^{5}$, which is well satisfied for the estimate $N\sim 10^{3}$, and thus such a system satisfies the necessary condition to be of Type I. 

The fact that the $(N)_{\text{max}}$ condition is satisfied but not the $(NV)_{\text{max}}$ condition additionally means that the system is located below the $T=\TII$ line in a plot like \reffig{2NVplot}, and thus the binding energies for the system are large enough that, at equilibrium, most of the TF-DNA bindings are correct (i.e., specific) bindings. Such a claim might seem strange given what is known about how TFs bind to DNA. TFs find their correct bindings sites through a two part process in which they first bind non-specifically to DNA and then slide along the DNA molecule. In the process of searching for its specific binding site, the TF spends most of its time non-specifically bound to DNA \cite{von1989facilitated}. This fact seems to contradict our claim that a TF-DNA system is dominated by specific rather than non-specific contacts. However, the TF's search for its correct binding site is a decidedly non-equilibrium process while our result is an equilibrium one. What our result suggests is that if the relaxation to equilibrium was not, for whatever reason, too slow for cellular function, TFs would still have sufficiently strong binding to their specific sites that they could successfully wade through the combinatorial sea of incorrect binding sites and find their correct ones. In other words, although real TF-DNA systems have evolved to not make use of equilibrium self-assembly, their biophysical properties appear to still afford them the ability to do so.

\subsection{Protein-Protein Interactions}

Although proteins are the ostensible conclusion of the central dogma of molecular biology, the basic unit of life is much more complex than a bag of freely diffusing proteins \cite{alberts1998cell}. Cells have highly organized internal structures with some proteins existing freely within the cramped environment of the cytoplasm while others function alongside organelles in complex-machine like interaction networks necessary for cellular metabolism or replication. But while a "bag of proteins" is not a faithful metaphor of the cell, it still serves as a useful model for studying the constraints of protein-protein interactions. 

Say we have a solution of $2N$ distinct monomeric proteins each of which, through a functional interaction, typically forms a heterodimer (and has the lowest binding energy) with one other protein, but also has the ability to bind to the other proteins through non-functional interactions (\reffig{protein_protein}). In terms of the dimer model, functional interactions correspond to correct dimers and non-functional interactions correspond incorrect dimers. Whether a non-gendered or a gendered dimer model is more appropriate when describing proteins depends on the interaction properties of the proteins involved. However, the two classes of models have sufficiently similar quantitative properties that we can choose the non-gendered model as representative of both. The reaction equation for such a (non-gendered) system would be 
\begin{equation}
\text{protein}_{k} + \text{protein}_{\ell} \xrightleftharpoons[ ]{}  \text{(protein-protein)}_{k\ell}
\end{equation}
where $k, \ell=1, \ldots, N$.

We consider again the question we asked for the previous biophysical systems: Given the approximate range of binding energies for protein dimers, are such protein-protein interactions systems Type I or Type II?

The authors of \cite{kastritis2011structure} provide a downloadable protein-protein interaction data set consisting of a collection of 144 protein complexes including antibody-inhibitor, enzyme-inhibitor, and G protein complexes. From this data set we can estimate an average binding free energy for functional protein complexes.  An estimate of the binding free energy for non-functional complexes is provided in  \cite{zhang2008constraints} by comparing the results of Yeast 2-Hybrid experiments across two data sets. Extracting our binding energy parameters $E_0$ and $\Delta$ from these data sets, we find $E_0 \simeq 18.9$ kcal/mol and $\Delta \simeq 7.7$ kcal/mol. We will take the mass of a monomer in this system to be the typical protein mass $m_{0} \simeq 64 \text{ kDa}$ \cite{milo2015cell}, and we assume a system temperature of $T = 310.15$ K.

With these parameter values, \rfw{search}  and \rfw{comb} give us, respectively,
\begin{equation}
NV < 4.7 \times 10^{-1} \text{ $\mu$m}^3\, , \qquad 2N < 2.7 \times 10^5, 
\end{equation} 
indicating that for a volume of $1\, \mu$m$^3$, the search-limiting constraint \rfw{search} provides a stronger limit on the number of different proteins in the system. Estimates of the number of different proteins in \textit{E. coli} put the number to be on order of $N \sim 10^3$ \cite{soufi2015characterization, corbin2003toward}, a result which satisfies the $(N)_{\text{max}}$ condition but not the $(NV)_{\text{max}}$ condition.  Given the calculated parameter values, we can check that $N\sim 10^3$ is more than three orders of magnitude less than the maximum computed from \rfw{2N_ineq}, and thus this system indeed satisfies the necessary condition to be of Type I. Therefore, like systems of interacting TFs and DNA sites, systems of interacting proteins in an \textit{E. coli} volume appear to be Type I (search-limited) dimer systems and thus have functional binding energies which are strong enough to overcome the combinatorial disadvantage of correct contacts at physiological temperatures. 

\renewcommand{\arraystretch}{1.2}
\begin{table*}[t]
\begin{ruledtabular}
\begin{tabular}{@{}lcccc|ccc@{}}
System & $m_0$ (kDa) & $E_0$ (kcal/mol)  &  $\Delta$ (kcal/mol) & $N_{\text{real}}$ & $(NV)_{\text{max}}$ ($\mu$m$^{3}$)   &    $(2N)_{\text{max}}$  & RHS of \rfw{2N_ineq} \\ \hline
ssDNA-ssDNA & $6.5$  & $0$  &  $31.5$  & $\sim 10^{12}$  & $\sim10^{4}$ &  $\sim 10^{22}$   & $\sim\exp(10^{13})$ \\
TF-DNA & $64$  & $22.9$   & $6.4$  & $\sim 10^{2}$  &   $\sim 1$ & $\sim 10^{4}$ & $\sim10^5$  \\
protein-protein & $64$ & $18.9$ &  $7.7$ & $\sim 10^{3}$ & $\sim 10^{-1}$  & $\sim10^5$ &  $\sim10^7$ \\
\end{tabular}
\end{ruledtabular}
\caption{\label{tab:table1}%
The energy and mass parameters and associated limits from \rfw{search}, \rfw{comb}, and \rfw{2N_ineq} for various biomolecular systems at $T = 310.15$ K. The ssDNA has 20 bases, and the $N_{\text{real}}$ values for TF-DNA and protein-protein are associated with \textit{E. coli}. Because TF-DNA interactions constitute a gendered dimer system, we used \rfw{search_1}, \rfw{comb_1}, and \rfw{N_ineq} to compute the relevant quantities in the TF-DNA row. In calculating \rfw{2N_ineq} (or \rfw{N_ineq}), we assumed a volume $V = 1\, \mu\text{m}^3$. The fourth column contains real upper limits on the monomer diversity of the associated systems. We see that although the values of $N_{\text{real}}$ exist below $(N)_{\text{max}}$ for each biomolecular system, $N_{\text{real}}$ exceeds $(NV)_{\text{max}}$ for a volume of $1\, \mu \text{m}^3$. Together, these two comparisons indicate that all of these systems are Type I (i.e., search-limited) dimer systems for a volume of $ 1\, \mu\text{m}^3$. Further affirming this label is that $N_{\text{real}}$ satisfies the Type I necessary condition \rfw{2N_ineq} for each system. Therefore, these biomolecular systems would have equilibrium curves for $\langle k \rangle$ and $\langle m\rangle$ more akin to those in \reffig{typeIplot} than to those in  \reffig{typeIandIIplot} or \reffig{typeIIplot}. }
\end{table*}

Actual protein-protein interaction systems have numerous features not present in the model.  Aside from the fact that proteins exist in multiple copies in real cells, we know that not all protein dimers are heterodimers (or even interact most strongly as heterodimers \cite{lukatsky2006statistically}); not all protein dimers can spontaneously dissociate into their constituent monomers (e.g., HIV-1 reverse transcriptase); not all constituent monomers are stable by themselves \cite{nooren2003diversity}; and not all proteins form dimers at all since many protein complexes (e.g., \textit{lac} repressor) contain more than two constituent proteins. 

But working within the constraints of the model, the fact that the estimated diversity of proteins in \textit{E. coli} is much lower than $(N)_{\text{max}}$ suggests that these protein systems have energy advantages for correct contacts that are larger than what would be marginally necessary to privilege those correct contacts in an equilibrium system. 

\section{Discussion and Interpretation \label{sec:discuss1}}

This work has five main analytical results: The exact partition function for dimer assembly, \rfw{final_part}; the associated equilibrium conditions,  \rfw{eqbm_k} and \rfw{eqbm_m}; the temperature condition for fully correct dimerization, \rfw{Tc}; the analytical definition of the two different system types, \rfw{sys}; the necessary but not sufficient inequalities for fully correct dimerization, \rfw{search} and \rfw{comb}; the necessary but not sufficient condition for the system to be of Type I, \rfw{2N_ineq}. 

The final two results allow us to qualitatively characterize two different system types. Contingent on a dimer system's binding energy, particle number, and volume parameters it can be categorized as Type I (search-limited), Type II (combinatorics-limited), or indeterminate. In search-limited systems, the energy advantage for correct contacts is large enough to overcome the combinatorial disadvantage of such contacts, and the achievement of the fully correct dimerization regime is more constrained by the ability of the correct monomers to find one another in their surrounding volume. In combinatorics-limited systems, the opposite is the case with binding energies being large enough for the monomers to find one another, and achieving fully correct dimerization is more constrained by the ability of the correct dimers to overcome their combinatorial disadvantage. Indeterminate systems have properties that cannot be cleanly distinguished as being either search-limited or combinatorics-limited. 

In terms of their binding trends, the qualitative difference between the two main types is that search-limited systems can be partially dimerized with most of their dimers consisting of correct contacts (\reffig{typeIplot}), whereas when combinatorics-limited systems are partially dimerized, most of the dimers consist of incorrect contacts (\reffig{typeIIplot}). Thus being able to categorize a dimer system as either Type I or II allows us to determine whether there can be mostly correct dimers in the system when the monomers are only partially dimerized.

Applying these results to the biophysical systems that motivated the model (\reffig{dimer_int}), we found that all such systems appear to be search-limited systems (Table \ref{tab:table1}). Per our previous discussion, this means that the fully correct dimerization regime in these systems is more constrained by the ability of monomers to find one another in their constituent volumes than by the need to overcome the combinatorial disadvantage of correct dimers, and that these systems are capable of having partially-dimerized states that are dominated by correct contacts. 

The latter result might appear obvious: Of course we should expect biomolecular systems with functional interactions to exhibit binding energies that privilege those functional interactions over competing ones. However, in most biophysical analyses of non-functional interactions (e.g., \cite{deeds2007robust, zhang2008constraints, johnson2011nonspecific}) emphasis is placed on how binding energies must be large enough to out compete non-functional interactions, and there is rarely any mention of how system size (in terms of volume) affects correct binding. But the interpretation behind the search-limiting condition \rfw{search} is that system size also constrains the ability of monomers to find one another and is just as relevant as binding energies in limiting non-functional interactions. 

This interpretation leads us to a second interesting result: \rfw{search} indicates that in achieving the fully correct dimerization regime, it is the product of particle number and volume, rather than their ratio encoded in density, that is constrained. This result reflects the fact that each of the monomers in a dimer system must find its optimal binding partner in the constituent volume, a task which is more difficult when said volume is large. This is because the quantity $2N$ serves two roles in this model; it defines the number of monomers in the system, but, since each monomer is distinct, it also defines the number of monomer species. Thus increasing $N$ increases the density of the system, leading to more interactions between monomers for a given volume, but it also increases the number of different interacting monomer types and makes it more difficult for a single monomer to find its one other optimal binding partner. Similarly, increasing the volume $V$ increases the number of position states a monomer must search through to find its optimal binding partner and makes such a search more difficult. Importantly, these effects are not independent. \rfw{search} indicates that the search-condition can be violated just as well for a large number of different monomers in a small volume as for a small number of monomers in a large volume. The "Dance Hall problem" discussed in \refsec{dance} is useful in lending an intuitive picture to the competing relevance of $N$ and $V$ in achieving fully correct dimerization: It is easiest for a person to reach his or her original dance partner if both the number of other dancers and the volume of the hall is small. Increase either one and the task of reaching one's partner becomes more difficult.

\section{Limitations and Extensions \label{sec:discuss2}}

To simplify our study of dimer self-assembly, we made a number of assumptions which limited the generality of the model and which thus point to ways to extend it.

First, we assumed that there was only a single-copy of each unique monomer in the system. This assumption greatly simplifies the combinatorial problem at the heart of the model, but does not match the properties of real biomolecular systems which always have many different monomer species each with a particular number of copies. However, one could consider a system where monomer species occur in multiple copies, but for which all monomers have the same copy-number. If these copies are uniformly distributed throughout the system, then for a small region, one can take the equilibrium dynamics of the system to be defined by the consideration of only a single copy of each species.

To move beyond such a heuristic argument would require a more general formulation of the problem. For example, the non-gendered model should include $2N$ unique monomers $\alpha_1, \ldots, \alpha_{2N}$ where an $\alpha_k$ monomer has $n_k$ copies in the system. For this more general system, one would need to determine the best way to model interactions between the same species and also how to consider mismatches between the number of possible correct partners and the number of available monomers in the system. Currently, it is not clear what is the best route towards attacking this more general problem.

For tractability, we did not give the monomers and dimers any sub-structure and instead defined their translational thermodynamics merely by the standard ideal-gas partition function \rfw{free}. But in protein systems, for example, we should expect the monomers and dimers to have non-zero moments of inertia and the dimers to have vibrational properties, properties we can incorporate into the preliminary partition function \rfw{sop_part} by correcting the quantities raised to the power of $k$ and $m$ with the appropriate rotational and vibrational partition functions. The principal effect of these contributions would be to give stronger temperature dependences to $\langle k \rangle$ and $\langle m \rangle$. For example, taking the monomers to be spherical and the dimers to be vibration-less linear molecules with moments of inertia $I$, the factor of  $\lambda_0^3/V\sim T^{-3/2}$ in \rfw{eqbm_k} would be replaced with 
\begin{equation}
\lambda_0^3 \Theta /V T \sim T^{-5/2},
\end{equation} 
where $\Theta = \hbar^2/2Ik_B$. It is apparent that for protein systems such incorporations are important because rotational and vibrational contributions to entropy have non-negligible contributions to the "price of lost freedom" \cite{finkelstein1989price} experienced by monomers when they associate into dimers. However, it is not clear whether these incorporations would remove the sharp fall off in $\langle m\rangle$ exhibited by Type II systems. 

Also, by giving the monomers and the dimers partition functions of the form $V/\lambda_0^3$, we assumed that they were dimensionless particles which did not interact outside of their bindings. Such an assumption is not correct for the aqueous, and often crowded, solutions in which biomolecules actually reside \cite{fulton1982crowded}. Thus, for better correspondence with real systems, we should incorporate volume exclusion and interparticle interactions into the model by replacing the ideal gas partition function \rfw{free} with the appropriate first-order terms in a Virial expansion \cite{kardar2007statistical_ch5}. 

We could more generally incorporate these caveats about the monomer and dimer partition functions by rewriting our analysis in terms of more general single-particle partition functions. Rather than taking monomer partition functions to be $V/\lambda_0^3$, we can take them to be $Q_{\text{m}}$, and rather than taking the dimer partition function to be proportional to $2 \sqrt{2} V/\lambda_0^3$, we can take the correct-dimer and incorrect-dimer partition functions to be $Q_{\text{d, corr}}$ and $Q_{\text{d, incorr}}$ respectively. Our partition function expression \rfw{sop_part}, would then become
\begin{align}
Z_{N}(V, T) & = \sum_{k=0}^N \sum_{m=0}^{k} \Omega_{N}(k, m)\,\left(\frac{Q_{\text{d, corr}}}{Q_{\text{d, incorr}}}\right)^{m}\mm
 & \qquad \qquad \times \left(Q_{\text{m}} \right)^{2N-2k}  \left(Q_{\text{d, incorr}} \right)^{k}.
\end{align}
And following the derivation in this paper, the associated equilibrium conditions for an average $\langle k \rangle$ dimers and $\langle m \rangle$ correct dimers is
\begin{align}
\frac{2 \,Q_{\text{d, incorr}}}{(Q_{\text{m}})^2}  & =  \frac{\langle k \rangle - \langle m \rangle (1 - \gamma^{-1})}{\big(N- \langle k \rangle\big)^2 } \mm[.75em]
\frac{\gamma}{2}  & = \langle m \rangle  \frac{N - \langle m \rangle (1- \gamma^{-1})}{\langle k \rangle - \langle m \rangle (1- \gamma^{-1})},
\end{align}
where we defined 
\begin{equation}
\gamma \equiv \frac{Q_{\text{d, corr}}}{Q_{\text{d, incorr}}}.
\end{equation}
Following the derivation further, we find the combinatorics-limited condition becomes 
\begin{equation}
 \frac{Q_{\text{d, corr}}}{Q_{\text{d, incorr}}} > 2N,
 \label{eq:comb_new}
\end{equation}
and the search-limited condition becomes
\begin{equation}
\frac{Q_{\text{d, corr}}}{(Q_{\text{m}})^2} > 2N,
\label{eq:search_new}
\end{equation}
both of which again depend on a finite-number of monomers in the system. \rfw{comb_new} and \rfw{search_new} are the more general analogs of the inequalities in \rfw{conds_sa} and thus they could be applied to dimer self-assembly systems which have more general partition functions than what is considered in the body of the paper.

Also in defining the binding energies of our dimer system, we took an ordinarily $2N\times 2N$ interaction matrix for the $2N$ distinct monomers to be characterized by only two parameters: $E_0$ and $\Delta$. Most generally, we would expect each species to have a specific interaction strength with every other species, thus requiring $N(2N-1)$ parameters. To work in this direction, we could consider an intermediate model with $N+1$ parameters where all incorrect interactions have binding energy $-E_0$ and the $i$th correct interaction has a binding energy $-(E_0  +\Delta_i)$, where $\Delta_i$ depends on the correct dimer species.

Two final limitations of the model concern length and time scales. Although the dimer model was able to capture some of the combinatorial properties of self-assembly, usually (as in the case of protein capsid or bilayer membrane assembly) the phrase "self-assembly" refers to the spontaneous construction of macromolecular structures that are much larger than their constituent parts \cite{israelachvili1976theory}. Thus, generalizations of this model that seek to provide more insight into the statistical physics constraints of self-assembly would need to incorporate self-assembly on a hierarchy of scales without sacrificing the precision of the statistical physics treatment. 

Second, since systems exhibiting self-assembly evolve towards equilibrium (rather than being perennially perched there), a mathematical model of the non-equilibrium properties of this dimer system would make a more useful archetype of self-assembly. Simulations are a good first step in this direction as long as they properly model the transition-state properties of assembly. To produce the simulations shown in \reffig{typeIandII}, we started all of our systems in the low-entropy microstate of all correct dimers and used a non-physical transition step in which dimers could switch constituent monomers without dissociating. These unphysical choices were meant to ensure that our system efficiently explored the state space over our chosen simulation times. However, a more faithful simulation of self-assembly would have the system begin in a state of all monomers and would only allow monomer dissociation and association as transition steps. Our preliminary attempts to abide by these constraints reveal that for certain parameter regimes the system falls prey to the common self-assembly problem of "kinetic traps" \cite{hagan2011mechanisms} in which even if the parameter space diagrams in \reffig{phase_dimer} suggest that the system is in the regime of fully correct dimerization, the system can remain, for long simulation times, in a state of only partially-correct dimers. This kinetic trapping appears to be most prevalent in Type II (combinatorics-limited) systems, and reasonably disappears as $E_0 \to 0$, suggesting the Type I vs. Type II categorization can also be a qualitative categorization for the likelihood of kinetic trapping. A more precise analytical argument would be preferred over these qualitative observations. 

\section{Conclusion \label{sec:conclusion}}

Motivated by the assembly of ssDNA into dsDNA, TF-DNA binding, and protein-protein interactions, we posed a statistical physics model in which a system of monomers could bind together in correct or incorrect dimers. We found that such systems could generally be "combinatorics-limited" or "search-limited" depending on how the energy parameters related to the number of monomers and the volume.  Combinatorics-limited systems have their correct bindings more constrained by the bindings overcoming their combinatorial disadvantage, and search-limited systems have their correct bindings more constrained by the bindings finding one another in the system. Real systems of dimer assembly (e.g., TF-DNA binding) all appear to be of the search-limited type meaning that their correct-energy contacts bind strongly enough to overcome the combinatorial cost of such contacts. Ultimately, the value in exploring such questions through a finite and analytical statistical physics framework rather than through the law of mass action or computation (which are more typical approaches) is that the finiteness and generality of an analytical partition function in statistical physics allows us to respect--and hence better account for--the finite-number combinatorial arrangements that are crucial in determining the possibility of assembly.

\section{Supplementary Material \label{sec:supp}}
Background calculations are included in the Supplementary Materials document. Jupyter code for creating \reffig{typeIandII}, \reffig{phase_dimer}, and for the biophysics calculations in \refsec{bio} can be found at: \\\href{https://github.com/mowillia/DimerSelfAssembly}{https://github.com/mowillia/DimerSelfAssembly}.

\section{Acknowledgments}
The author thanks Sanchari Bhattacharyya, Michael Manhart, and Seong-Ho Pahng for helpful conversations, Vinny Manoharan, Eugene Serebryany, and Eugene Shakhnovich for suggestions about applicable biomolecular systems, and Rostam Razban for reading, editing, and discussing drafts of the work. This work was funded by the Harvard Graduate Prize Fellowship.

\appendix{}

\section{Gendered System \label{app:gen}}

In \refsec{ngpt}, we introduced our study of the self-assembly of a dimer system by considering a collection of monomers where each monomer could form a dimer with any other monomer. In this sense, we labeled this system as "non-gendered" to differentiate it from systems in which monomers have constraints on the type of monomers to which they can bind. In this section, we introduce a model with such constraints, namely one in which there are two types of monomers and each monomer can only form a dimer with the monomer of the opposite type. The statistical physics analysis of this gendered dimer system is very similar to that of the non-gendered system, so we focus on the major results rather than derivations. 

\subsection{Gendered partition function}

Say that our system contains $2N$ distinguishable monomers of two kinds. There are $N$ distinguishable monomers labeled $\beta_1, \beta_2, \ldots, \beta_N$ each of which has mass $m_{\beta}$, and there are $N$ distinguishable monomers labeled $\alpha_1, \alpha_2, \ldots, \alpha_N$ each of which has mass $m_{\alpha}$. The $2N$ total monomers exist in thermal equilibrium at temperature $T$ and in a volume $V$. Each $\alpha$ monomer can bind to any $\beta$ monomer (and vice versa), but $\alpha$ monomers cannot bind to each other, and $\beta$ monomers cannot bind to each other. When monomer $\alpha_k$ binds to monomer $\beta_{\ell}$, the two form the dimer $(\alpha_k, \beta_{\ell})$, where the ordering within the pair is not important. We define correct dimers as those consisting of $\alpha_{k}$ binding to $\beta_{k}$ for $k=1, \ldots, N$; all other dimers are considered incorrect. Thus there are $N$ possible correct dimers in this system and $N(N-1)$ possible incorrect dimers. The binding energy for the dimers is given by 
\begin{equation}
{\cal E}'(\alpha_m, \beta_n) = \begin{dcases}- (E_0 + \Delta) & \text{ if $m = n$} \\ - E_0 & \text{ if $m \neq n$,}\end{dcases}
\end{equation}
indicating that correct dimers have a binding energy of $-(E_0+\Delta)$ and incorrect dimers have a binding energy of $-E_0$, where $E_0, \Delta >0$. 

\begin{figure}[t]
\centering
\includegraphics[width=.85\linewidth]{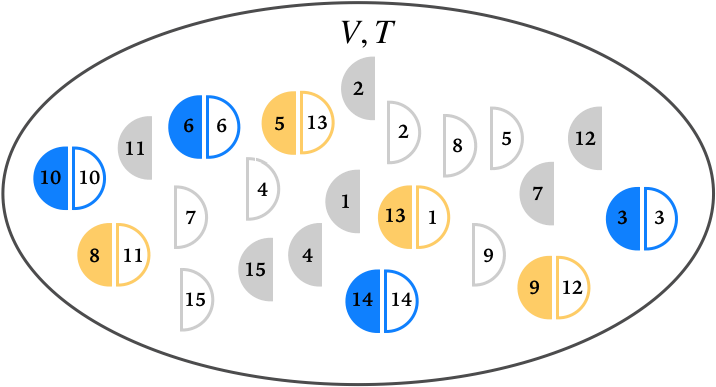}
	\caption{(Color online) Example microstate of the gendered system with $2N=30$ subunits. We represent the monomers of either gender as filled or unfilled half circles. Filled half-circles can only bind to unfilled half-circles. Correct dimers consist of binding $k$ to $k$ and have binding energy $-(E_0+\Delta)$. All other dimers are incorrect and have binding energy $-E_0$. This microstate has four correct dimers (in blue), four incorrect dimers (in yellow), and fourteen monomers (in grey). The total binding energy for this microstate is $-(8 E_0 + 4 \Delta)$. For pictorial clarity, the figure represents monomers as half-circles, but monomers are taken to be point particles in the model.  }
	\label{fig:gen_sys}
\end{figure}

We assume that the monomers and dimers are point particles with no rotational or vibrational properties and that apart from the binding energy, the monomers and the dimers are free particles that do not interact with one another. An example microstate for this system is shown in \reffig{gen_sys}. 

We want to compute the partition function for this system. By an argument similar to that used to establish \rfw{sop_part} and \rfw{omeg}, we find that the partition function can be written as 
\begin{align}
Z'_{N}& (V, T, E_0 , \Delta) \mm
& =  \quad \sum_{j=0}^{N}\sum_{\ell=0}^{j} \binom{N}{\ell} b_{N-\ell, j-\ell}\,e^{\beta(j  E_0 + \ell \Delta)}\mm
& \quad \times  \left(\frac{V}{\lambda_{\alpha}^3}\right)^{N-j} \left( \frac{V}{\lambda_{\beta}^3}\right)^{N-j} \left(\frac{V}{\lambda_{\alpha \beta}^3}\right)^{j}
\label{eq:sop_part2gen}, 
\end{align}
where $\lambda_{\alpha}$, $\lambda_{\beta}$, and $\lambda_{\alpha \beta}$ are the thermal de Broglie wavelengths of an $\alpha$ monomer, a $\beta$ monomer, and an $(\alpha, \beta)$ dimer respectively. In the summations in \rfw{sop_part2gen}, $j$  counts the number of dimers in the system, and $\ell$ counts the number of correct dimers. The factor 
\begin{equation}
 \binom{N}{\ell} b_{N-\ell, j-\ell}
 \label{eq:bnm}
\end{equation}
is the answer to the following question:
\begin{quote}
$N$ man-woman pairs enter a dance hall. All the pairs separate, and people mingle with one another such that at some later time, there are some man-woman pairs and there are some men and women who are alone. At this later time, there are $j$ man-woman pairs on the dance floor, and of this set, there are $\ell$ pairs from the set of original pairs. How many ways can this happen? 
\end{quote}

Interpreting \rfw{bnm}  more physically, the factor $\binom{N}{\ell}$ corresponds to the number of ways to choose $\ell$ dimers from the set of $N$ possible correct dimers. Under the constraint that each dimer consists of opposite gender monomers, the factor $b_{N-\ell, j-\ell}$ is the number of ways of forming $j-\ell$  dimers from a set of $2(N-\ell)$ monomers such that none of the chosen dimers is amongst the set of $N-\ell$ correct dimers. 

In computing \rfw{sop_part2gen}, the pivotal quantity is $b_{N-\ell, j-\ell}$. We can determine this quantity by considering another question:
\begin{quote} Given $n$ original man-woman pairs, what is the number of ways to form $k\leq n$ man-woman pairs such that none of these new pairs coincide with any of the original pairs?  
\end{quote}
We call this number $b_{n, k}$. Applying the principle of inclusion and exclusion in a way similar to the application in \textit{SM Sec. C}, we find 
\begin{equation}
b_{n, k} = \sum_{m=0}^{k} (-1)^{m} \binom{n}{m} \binom{n-m}{k-m}^2 (k-m)!
\end{equation}
Using the definition of the Gamma function to express $(n-m)!$ as an integral, we then obtain
\begin{equation}
b_{n, k} = \frac{1}{(n-k)!} \binom{n}{k} \int^{\infty}_{0} dx\, e^{-x} \, x^{n-k} (x-1)^{k}.
\label{eq:bnl_intg}
\end{equation}
As a consistency check, we can use \rfw{bnl_intg} to prove the identity 
\begin{equation}
\binom{N}{j}^2 j ! = \sum_{\ell=0}^{j} \binom{N}{\ell} b_{N-\ell, j-\ell},
\label{eq:bnl_sum_id}
\end{equation}
which asserts that the total number of unique ways to form $j \leq N$ man-woman pairs (regardless of coincidence with some original pairing), is the  number of ways to choose $\ell$ original pairs multiplied by the number of ways to choose $j-\ell$ non-original pairs and summed over $\ell$. The method of proof for \rfw{bnl_sum_id} is similar to that applied at the end of \textit{SM Sec. C}.

We are now ready to return to \rfw{sop_part2gen}. First, we rewrite the translational partition function contributions to the main partition function as 
\begin{equation}
\left(\frac{V}{\lambda_{\alpha}^3}\right)^{N-j} \left( \frac{V}{\lambda_{\beta}^3}\right)^{N-j} \left(\frac{V}{\lambda_{\alpha \beta}^3}\right)^{j} = \left( \frac{V}{\bar{\lambda}^3}\right)^{2N}  \left( \frac{\lambda_{\mu}^3}{V} \right)^{j}, 
\end{equation}
where we defined
\begin{equation}
\bar{\lambda} \equiv \frac{h}{\sqrt{2\pi (m_{\alpha} m_{\beta})^{1/2} k_{B}T}}, \quad \lambda_{\mu} \equiv \frac{h}{\sqrt{2\pi k_BT}} \sqrt{\frac{1}{m_{\alpha}} + \frac{1}{m_{\beta}}}, 
\label{eq:bid}
\end{equation} 
Now, with the Laplace's integral form of the Legendre Polynomial $P_{n}(x)$ \cite{gould2010ver5}
\begin{equation}
P_{n}(x) = \frac{1}{2\pi} \int^{2\pi}_{0} d\phi \left( x + \sqrt{x^2-1} \cos \phi\right)^{n}. 
\end{equation}
and the series representation of the Legendre Polynomial \cite{gould2010ver5}
\begin{equation}
P_{n}(x) = \left( \frac{x-1}{2} \right)^{n} \sum_{k=0}^n \binom{n}{k}^2 \left(\frac{x+1}{x-1}\right)^{k}.
\end{equation}
we can establish the integration identity
\begin{equation}
\sum_{k=0}^{n} \binom{n}{k}^2 u^k = \frac{1}{2\pi} \int^{2\pi}_{0} d\phi\, \left( 1 + u + 2\sqrt{u}\cos\phi\right)^{n}.
\label{eq:intg_id}
\end{equation}

Incorporating \rfw{bnl_intg} into \rfw{sop_part2gen}, following a derivation analogous to that in \textit{SM Sec. D}, and using \rfw{intg_id}, we ultimately find that the partition function for this system is 
\begin{widetext}
\begin{equation}
Z'_{N}(V, T, E_0, \Delta) = \frac{1}{2\pi N!} \left( \frac{V}{\bar{\lambda}^3}\right)^{2N}\int^{2\pi}_{0} d\phi\, \int^{\infty}_{0} \int^{\infty}_{0}dx \,dy \, e^{-(x+y)} \,{\cal I}^{N},
\label{eq:ztfin}
\end{equation}
where
\begin{align}
{\cal I} \equiv x + \frac{\lambda_{\mu}^3}{V} \, e^{\beta E_0} \,y\, \Omega(x; \beta \Delta) - 2 \left(\frac{\lambda_{\mu}^3}{V} \right)^{1/2} e^{\beta E_0/2} \sqrt{y \,x \,\Omega(x; \beta \Delta)}\, \cos \phi, 
\label{eq:Idef}
\end{align}
\end{widetext}
and 
\begin{equation}
\Omega(x; \beta \Delta) \equiv e^{\beta \Delta} + x -1.
\end{equation}
The thermal de Broglie wavelength in these expressions is defined as $\lambda_{\mu} = h/\sqrt{2\pi \mu k_BT}$ with $\mu = m_{\beta}m_{\alpha}/(m_{\beta} + m_{\alpha})$, the reduced mass of an $(\alpha,\beta)$ dimer.

\subsection{Equilibrium Conditions}
With \rfw{ztfin}, the next step in studying the equilibrium properties of the gendered dimer system is to derive the equilibrium conditions. Given \rfw{sop_part2gen}, we see that we can compute the average number of total dimers and the average number of correct dimers, respectively, with 
\begin{align}
\langle j \rangle & = \frac{\partial}{\partial (\beta E_0)} \ln Z'_{N}, \label{eq:jandldef_1}\\
\langle \ell \rangle & = \frac{\partial}{\partial (\beta \Delta)} \ln Z'_{N}.
\label{eq:jandldef_2}
\end{align}
We can also compute the variances and covariances between these quantities through
\begin{equation}
\left( \begin{array}{c c} \sigma_{j}^2 & \sigma_{j \ell}^2 \\[0.5em] \sigma_{ \ell j}^2 & \sigma_{\ell}^2 \end{array}\right) = \left( \begin{array}{c c} \partial^2_{\beta E_0} & \partial_{\beta E_0}\partial_{\beta \Delta} \\[0.5em]  \partial_{\beta \Delta}\partial_{\beta E_0} & \partial^2_{\beta \Delta}\end{array}\right) \ln Z'_{N},
\label{eq:cov_mat_ellj}
\end{equation}
where  $\sigma_{j}^2$ is the variance in the total number of dimers, $\sigma_{\ell}^2$ is the variance in the number of correct dimers, and $\sigma_{\ell j}^2 = \sigma_{j \ell}^2$ is the covariance between the total number of dimers and the number of correct dimers.

Using \rfw{ztfin} directly in \rfw{jandldef_1} and \rfw{jandldef_2} would result in cumbersome integral expressions for $\langle \ell \rangle$ and $\langle j \rangle$, so we will use Laplace's method to approximate the partition function. We can expect the exact calculation of this approximation to mirror that in  \textit{SM Sec. E}, but first we need to reduce \rfw{ztfin} from a three-dimensional to a two-dimensional integral. Implementing Laplace's method on the $\phi$ variable alone, we find that the integrand of \rfw{ztfin} is maximized for $\phi= \pi$. Therefore, we can make the approximation 
\begin{align}
\ln Z'_{N}(V, T, E_0, \Delta) & = \ln \int^{\infty}_{0} \int^{\infty}_{0}dx \,dy \, e^{-(x+y)} \,{\cal I}_{\phi = \pi}^N + \cdots,
\end{align}
where ${\cal I}_{\phi = \pi}$ is \rfw{Idef} evaluated at $\phi = \pi$ and where "$\cdots$" stands in for terms that are independent of $E_0$ and $\Delta$ or are sub-leading to order $N$. Now,  using \rfw{jandldef_1} and \rfw{jandldef_2} and implementing the standard Laplace's method algorithm in a way akin to its application in  \textit{SM Sec. E}, we find the system of equations
\begin{align}
\frac{\lambda_{\mu}^3}{V}\, e^{\beta E_0} &=  \frac{\langle j \rangle - \langle \ell \rangle (1 - e^{-\beta \Delta})}{\big(N- \langle j \rangle\big)^2 }, \label{eq:eqbm_k2}\\
 e^{\beta \Delta} &= \langle \ell \rangle  \frac{N - \langle \ell \rangle (1- e^{-\beta \Delta})}{\langle j \rangle - \langle \ell \rangle (1- e^{-\beta \Delta})}.
\label{eq:eqbm_m2}
\end{align}
We similarly find the variances and covariances between the number of dimers and the number of correct dimers are 
\begin{align}
\sigma_{j}^2 & = \frac{1}{2N} \langle j \rangle \big(N - \langle j \rangle\big), \\  
 \sigma_{ j\ell}^2 & =  \frac{1}{2N} \langle \ell \rangle \big( N - \langle j \rangle \big),\\
 \sigma_{\ell}^2 & = \langle \ell \rangle - \frac{\langle \ell \rangle^2}{2} \left( \frac{1}{\langle j \rangle} + \frac{1}{N} \right).
\end{align}
Comparing \rfw{eqbm_k2} and \rfw{eqbm_m2} with \rfw{eqbm_k} and \rfw{eqbm_m}, we see that the sets of equilibrium conditions for the non-gendered and gendered systems are identical except for numerical factors. Therefore, the discussion in the main text also applies to this gendered system with only slight changes to the arguments of important expressions. In particular, considering the fully correct dimerization condition for the gendered system  (i.e., $\langle j \rangle = \langle \ell \rangle$), we find that the critical temperature $k_BT_c = \beta_c^{-1}$ at which this condition is satisfied is 
\begin{equation}
\frac{\lambda_{\mu,c}^3}{V} \,e^{\beta_c(E_0 + \Delta)} \frac{\left(1- N e^{-\beta_c \Delta}\right)^2}{1-e^{-\beta_c\Delta}} = N-1,
\label{eq:Tc2}
\end{equation}
where $\lambda_{\mu, c} = h/\sqrt{2\pi \mu k_BT_c}$. 
Similarly to \rfw{sys}, we can categorize the system as Type I or II according to the limiting behavior of the solution to \rfw{Tc2}. We define $\TI'$ as 
\begin{equation}
k_B\TI' \equiv \frac{2}{3}(E_0 + \Delta) \left[W_{0} \left(\frac{2 (E_0 + \Delta)}{3 E_{\mu, V}} {N}^{2/3}\right)\right]^{-1}+{\cal O}\left( N^{-1} \right) 
\label{eq:T12}
\end{equation}
where $E_{\mu, V} \equiv h^2/2\pi \mu V^{2/3}$, and $\TII'$ as 
\begin{equation}
k_B\TII'\equiv \frac{\Delta}{\ln (N)}.
\label{eq:T22}
\end{equation}
Then a gendered system is Type I or Type II according to 
\begin{equation}
\text{System Type} = \begin{dcases} \text{Type I} & \text{ for $T_c \simeq \TI' $,}\\ \text{Type II} & \text{ for $T_c \simeq \TII'$.} \end{dcases}
\label{eq:sys2}
\end{equation}
The parameter space behavior of this system is identical to that in \reffig{phase_dimer}, with $\TI'$ and $\TII'$ replacing $\TI$ and $\TII$, respectively. 

\subsection{Inequalities for Assembly and Type}

With \rfw{T12} and \rfw{T22}, we can derive inequalities analogous to \rfw{search}, \rfw{comb}, and \rfw{2N_ineq}. 

For the gendered dimer system, the "search-limiting" condition, derived from $T< \TI'$, is 
\begin{equation}
NV < \lambda_{\mu}^3 \, e^{\beta (E_0+\Delta)},
\label{eq:search_1}
\end{equation}
where, consistent with the $N\gg1$ limit, we dropped the ${\cal O}(N^{-1})$ term in \rfw{T12}. 
The "combinatorics-limiting" condition, derived from $T< \TII'$, is 
\begin{equation}
N < e^{\beta \Delta}.
\label{eq:comb_1}
\end{equation}
\rfw{search_1} and \rfw{comb_1} are the two necessary, but not sufficient, conditions a gendered dimer system must satisfy to be in the fully correct dimerization regime of its parameter space. 

For a Type I dimer system, we require $\TI'< \TII'$. Using \rfw{T12} and \rfw{T22} in the inequality $\TI'< \TII'$, and noting that if $W_0(X)>k$, then $X> k e^{k}$, we obtain an inequality that when solved for $N$ yields
\begin{equation}
N < \exp\left[\frac{3\Delta}{2E_0}W_{0} \left(\frac{2E_0}{3E_{\mu,V}}\right)\right]
\label{eq:N_ineq}
\end{equation}
\rfw{N_ineq} is a necessary, but not sufficient, condition for a gendered dimer system to be of Type I.

\subsection{One type of monomer fixed; $m_{\alpha} \to \infty$ limit}

A special case of the gendered dimer system occurs when one of the two types of monomers is fixed in space. We can envision such a system as having $N$ distinguishable monomers interacting with $N$ binding sites where each monomer has a preferred binding site to which it binds with energy $-(E_0+\Delta)$; for all other binding sites, the monomer binds with energy $-E_0$.

\begin{figure}[t]
\centering
\includegraphics[width=.85\linewidth]{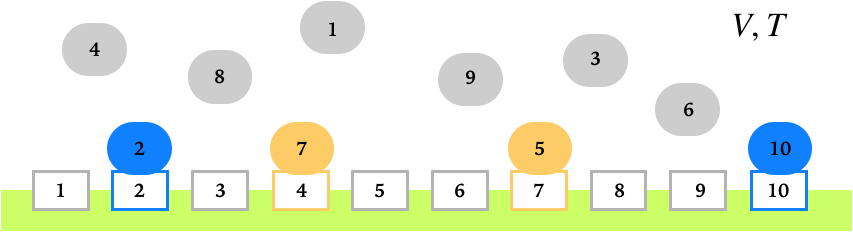}
	\caption{(Color online) Example microstate of the gendered system with $2N=20$ subunits where one type of monomer is fixed in space. We represent the two genders as shaded or unshaded shapes. This microstate has two correct contacts (in blue), two incorrect contacts (in yellow), and six monomers and unpaired binding sites (in grey). The total binding energy for this microstate is $-(4 E_0 + 2 \Delta)$. }
	\label{fig:fixed_gen_sys}
\end{figure}

An example microstate of such a system is shown in \reffig{fixed_gen_sys}. The general partition function for this system can be directly obtained from \rfw{ztfin} by removing the $V^N/\lambda_{\alpha}^{3N}$ factor from the coefficient and taking $\lambda_{\mu} \to \lambda_{\beta}$. That is, if we are taking the particles of type $\alpha$ to be fixed, then we ignore their dynamics by taking $m_{\alpha} \to \infty$, thus taking the reduced mass $\mu$ to $m_{\beta}$. 

The equilibrium conditions for this system are similarly given by \rfw{eqbm_k2} and \rfw{eqbm_m2} with $\lambda_{\mu}$ replaced with $\lambda_{\beta}$ in the former. 

\pagebreak

\onecolumngrid

\setlength\parskip{0in}

\fontsize{10}{11}\selectfont

\titleformat{\section}
  {\normalfont\fontsize{14}{12}\bfseries}{\thesection}{1em}{}
  
  \titleformat{\subsection}
  {\normalfont\fontsize{12}{12}\bfseries}{\thesubsection}{1em}{}

\section*{Supplemental Materials for "Self-Assembly of a Dimer System"}

\setcounter{equation}{0}
\setcounter{figure}{0}
\setcounter{section}{0}
\setcounter{table}{0}
\setcounter{page}{1}
\makeatletter
\renewcommand{\thefigure}{S\arabic{figure}}
\renewcommand{\bibnumfmt}[1]{[S#1]}
\renewcommand{\citenumfont}[1]{S#1}


\section{Link to Supplementary Code}

Jupyter notebooks for creating \reffig{typeIandII}, \reffig{phase_dimer}, and for the biophysics calculations in \refsec{bio} in the main text can be found at \href{https://github.com/mowillia/DimerSelfAssembly}{https://github.com/mowillia/DimerSelfAssembly}.


\section{Deriving $a_{n, \ell}$ as a Series and an Integral \label{app:anl}}

We are seeking a formula that answers the following question: 
\begin{quote} Given $2n$ distinguishable objects that are all initially paired in some way, what is the number of ways to form $\ell$ pairs such that none of these new pairs coincides with any original pairings?
\end{quote} 
We call this number $a_{n, \ell}$, and it is easy to see what its value should be for $\ell=n$ and $\ell =1$. If we were to take $\ell = n$, we would have the case of the "bridge couples problem" and we should obtain the formula derived in \cite{margolius2001avoidings1}. If were were to take $\ell=1$, we could infer that $a_{n, 1} = 2n(2n-2)/2$ since there are $2n$ ways to select the first element, $2n-2$ ways to select an element that was not initially paired with this first element, and a factor of $1/2$ for the fact that the order of this selection is not important. 

To find the general formula for $a_{n, \ell}$, we employ the inclusion-exclusion principle \cite{chuan1992principless1}. First, we establish some definitions. We define $|A_i|_{n,\ell}$ as the number of way to reform $\ell$ pairs, out of $2n$ initially paired elements, such that in the new set of pairs, we include the $i$th pair of the initial pairings. We in turn say that the quantity 
\begin{equation}
|A_{i_1} \cap \cdots \cap A_{i_k}|_{n, \ell}, 
\end{equation} 
equals the size of the set where, out of $2n$ initially paired elements, we have formed $\ell\leq n$ new pairs which include the pairs $i_1, \ldots, i_k$ (for $k \leq \ell)$ of the original pairings. By this definition, our desired quantity $a_{n, \ell}$ can be written as 
\begin{equation}
\sum_{1\leq i_1 < \cdots < i_{\ell} \leq n}^n |A^c_{i_1} \cap \cdots \cap A^c_{i_\ell}|_{n, \ell},
\label{eq:Acell}
\end{equation}
where $A^c_k$ is the complement of $A_k$. \refew{Acell} is the total number of ways to reform $\ell$ pairs out of $2n$ initially paired elements such that none of the $\ell$ pairs is found in the initial pairings. Given that the intersection of complements is equal to the complement of the union, we have. 
\begin{align}
\sum_{1\leq i_1 < \cdots < i_{\ell} \leq n}^n |A^c_{i_1} \cap \cdots \cap A^c_{i_\ell}|_{n,\ell} &  = \sum_{1\leq i_1 < \cdots < i_{\ell} \leq n}^n|\left(A_{i_1} \cup \cdots \cup A_{i_\ell} \right)^c|_{n, \ell}  = |{\cal S}|_{n,\ell} - \sum_{1\leq i_1 < \cdots < i_{\ell} \leq n}^n|A_{i_1} \cup \cdots \cup A_{i_\ell}|_n,
\label{eq:Acell2}
\end{align}
where we we defined $ |{\cal S}|_{n,\ell}$ as the number of ways to create $\ell \leq n$ pairs out of a set of $2n$ elements. Combinatorics tells us that $ |{\cal S}|_{n,\ell}$ is 
\begin{equation}
 |{\cal S}|_{n,\ell}= \binom{2n}{2\ell} \frac{(2\ell)!}{2^{\ell} \ell!} = \binom{2n}{2\ell} (2\ell-1)!!,
 \label{eq:snl}
\end{equation}

Now, to compute \refew{Acell}, we must calculate the last quantity in \refew{Acell2}, and we do so by the inclusion-exclusion principle. By the principle, we have
\begin{align}
\sum_{1\leq i_1 < \cdots < i_{\ell} \leq n}^n|A_{i_1} \cup \cdots \cup A_{i_\ell}|_{n,\ell}& =  \sum_{i=1}^{n}|A_{i}|_{n,\ell}  - \sum_{1\leq i < j \leq n}^n|A_{i} \cap A_{j}|_{n,\ell}\mm & 
\quad \quad+ \cdots + \sum_{1\leq i_1 < \cdots < i_{\ell} \leq n}^n(-1)^{\ell-1} |A_{i_1} \cap \cdots \cap A_{i_{\ell}}|_{n, \ell}.
\label{eq:incl_excl}
\end{align}
We recall that $|A_i|_{n,\ell}$ equals the number of way to reform $\ell$ pairs, out of $2n$ initially paired elements, such that in the new set of pairs, we include the $i$th pair of the initial pairings. Since the $i$th pair is fixed in this pairing, the number of ways to achieve this new pairing is simply the number of ways to form $\ell-1$ pairs out of a set of $2n-2$ elements. Thus we have 
\begin{equation}
|A_i|_{n,\ell} = \binom{2n-2}{2\ell-2} (2\ell-2-1)!!.
\end{equation}
This quantity is independent of which $i$ we choose, so, in \refew{incl_excl}, the summation can be replaced with the factor $\binom{n}{1}$. Similarly, the quantity $|A_i \cap A_j|_{n,\ell}$ is the number of ways to choose $\ell$ pairs, out of $2n$ initially paired elements, such that we include the $i$th and $j$th pairs of the original pairing. Thus, we have 
\begin{equation}
|A_i \cap A_j|_{n,\ell} = \binom{2n-4}{2\ell-4} (2\ell-4-1)!!,
\end{equation}
and the summation is replaced with the factor $\binom{n}{2}$. Following this pattern, we find that \refew{incl_excl} becomes
\begin{align}
\sum_{1\leq i_1 < \cdots < i_{\ell} \leq n}^n|A_{i_1} \cup \cdots& \cup A_{i_\ell}|_{n,\ell}  = \sum_{j=1}^{\ell}(-1)^{j-1} \binom{n}{j} \binom{2n-2j}{2\ell-2j} (2\ell-2j-1)!!. 
\end{align}
Finally, using \refew{snl} in \refew{Acell2}, and noting that the final result is our desired $a_{n, \ell}$, we have 
\begin{equation}
a_{n, \ell} = \sum_{j=0}^{\ell}(-1)^{j} \binom{n}{j} \binom{2n-2j}{2\ell-2j} (2\ell-2j-1)!!. 
\label{eq:anl0}
\end{equation}

We can also write \refew{anl0} as an integral which will later allow us to write the partition function as a double integral. The first step is to rewrite the second combinatorial factor as 
\begin{align}
\binom{2n-2j}{2\ell-2j} = \frac{2^{n-j} (n-j)!}{(2n-2\ell)!(2\ell- 2j)!} (2n-2j-1)!!. 
\end{align}
We then find 
\begin{align}
a_{n, \ell} & = \sum_{j=0}^{\ell} (-1)^{j} \binom{n}{j} \frac{2^{n-j} (n-j)!}{(2n-2\ell)!(2\ell- 2j)!}\, (2n-2j-1)!! \,(2\ell-2j-1)!!\mm
& =  \frac{2^{n-\ell}}{(2n-2\ell)!}  \frac{n!}{\ell!}  \sum_{j=0}^{\ell} (-1)^{j} \frac{\ell!}{j! (\ell-j)!} (2n-2j-1)!!  \mm
& = \frac{2^{n-\ell}(n-\ell)!}{(2n-2\ell)!}  \frac{n!}{\ell!(n-\ell)!}  \sum_{j=0}^{\ell} (-1)^{j} \binom{\,\ell\,}{j} \frac{2^{n-j}}{\sqrt{\pi}} \Gamma(n-j + 1/2).
\end{align}
Finally, using the integral definition of the Gamma function we obtain. 
\begin{equation}
a_{n, \ell}  = \frac{1}{(2n-2\ell-1)!!} \binom{n}{\ell} \frac{1}{\sqrt{\pi}} \int^{\infty}_{0} dt\, e^{-t} t^{-1/2} (2t)^{n} (1-1/2t)^{\ell}.
\label{eq:anl_i}
\end{equation}
For completeness, we can now use \rfw{anl_i} to check the condition \rfw{omeg2}. Computing the relevant quantity we have
\begin{align}
 \sum_{m=0}^{k}\binom{N}{m} a_{N-m, k-m} & =  \frac{1}{(2N-2k-1)!!}   \sum_{m=0}^{k}\binom{N}{m}\binom{N-m}{k-m} \frac{1}{\sqrt{\pi}} \int^{\infty}_{0} dt\, e^{-t} t^{-1/2} (2t)^{N-m} (1-1/2t)^{k-m}\mm
  & = \frac{1}{(2N-2k-1)!!} \frac{1}{\sqrt{\pi}} \int^{\infty}_{0} dt\, e^{-t} t^{-1/2} (2t)^{N}(1-1/2t)^{k}  \sum_{m=0}^{k}\binom{N}{m}\binom{N-m}{k-m}(2t-1)^{-m}.
\end{align}
With the combinatorial identity 
\begin{equation}
\binom{N}{m}\binom{N-m}{k-m} =  \binom{N}{k}\binom{k}{m},
\end{equation}
we have 
\begin{align}
 \sum_{m=0}^{k}\binom{N}{m} a_{N-m, k-m} & = \frac{1}{(2N-2k-1)!!} \binom{N}{k} \frac{1}{\sqrt{\pi}} \int^{\infty}_{0} dt\, e^{-t} t^{-1/2} (2t)^{N}(1-1/2t)^{k}  \sum_{m=0}^{k}\binom{k}{m}(2t-1)^{-m}\mm
  & = \frac{1}{(2N-2k-1)!!} \binom{N}{k} \frac{1}{\sqrt{\pi}} \int^{\infty}_{0} dt\, e^{-t} t^{-1/2} (2t)^{N}(1-1/2t)^{k} \left(\frac{2t}{2t-1} \right)^{k} \mm
   & =  \frac{1}{(2N-2k-1)!!} \binom{N}{k} \frac{1}{\sqrt{\pi}} \int^{\infty}_{0} dt\, e^{-t} t^{-1/2} (2t)^{N} \mm
   & = \frac{ (2N-1)!!}{(2N-2k-1)!!}\binom{N}{k},
\end{align}
where we used the identity 
\begin{equation}
\frac{1}{\sqrt{\pi}} \int^{\infty}_{0} dt\, e^{-t} t^{-1/2} (2t)^{N}  = \frac{2^N}{\sqrt{\pi}} \Gamma(N+1/2) = (2N-1)!!. 
\end{equation}
Further reducing the final expression gives us 
\begin{align}
 \sum_{m=0}^{k}\binom{N}{m} a_{N-m, k-m} & =  \frac{ (2N-1)!!}{(2N-2k-1)!!}\binom{N}{k}\mm
  & = \frac{(2N)!}{2^N N!} \frac{(N-k)!2^{N-k}}{(2(N-k))!} \frac{N!}{k!(N-k)!}\mm
   & = \frac{(2N)!}{(2k)! (2(N-k))!} \frac{(2k)!}{2^k k!} = \binom{2N}{2k} (2k-1)!!, 
 \end{align}
as expected from \rfw{omeg2}.
\section{Derivation of Non-Gendered Partition Function \label{app:ngpt}}

In deriving the final form of the partition function for the non-gendered system, we begin with the partition function expressed as a summation over the total number of dimers and the total number of correct dimers:
\begin{equation}
Z_{N}(V, T, E_0, \Delta) = \sum_{k=0}^N \sum_{m=0}^{k} \binom{N}{m} a_{N-m, k-m}\,e^{\beta (kE_0 +m\Delta)}\left(\frac{V}{\lambda_0^3} \right)^{2N-2k}  \left(\frac{V}{(\lambda_0/\sqrt{2})^3} \right)^k  \, 
\label{eq:sop_part2}
\end{equation}
Using the integral expression \refew{anl_i}, we find \refew{sop_part2} becomes 
\begin{align}
Z_{N}(V, T, E_0, \Delta)  & = \frac{(V/\lambda_0^3)^{2N}}{\sqrt{\pi}} \int^{\infty}_{0} dx\, \frac{e^{-x}}{\sqrt{x}}\sum_{k=0}^N \sum_{m=0}^k \delta^k \eta^m \frac{1}{(2N-2k-1)!!}  \binom{N}{m} \binom{N-m}{k-m} (2x)^{N-k} (2x-1)^{k-m}\mm
& = \frac{(V/\lambda_0^3)^{2N}}{\sqrt{\pi}}\int^{\infty}_{0} dx\, \frac{e^{-x}}{\sqrt{x}}(2x)^N\sum_{k=0}^N \sum_{m=0}^k \frac{\left[\delta(1- 1/2x)\right]^k}{(2N-2k-1)!!} \binom{N}{N-m} \binom{N-m}{k-m}\left( \frac{\eta}{2x-1} \right)^m
\label{eq:part2}
\end{align}
where we defined
\begin{equation}
\delta \equiv  \frac{ 2 \sqrt{2}\lambda_0^3}{V}e^{\beta E_0}, \qquad \eta \equiv e^{\beta \Delta}.
\label{eq:param_def}
\end{equation}
Next, we isolate the sum over $m$ to find
\begin{align}
\sum_{m=0}^k \binom{N}{N-m} \binom{N-m}{k-m}\left( \frac{\eta}{2x-1} \right)^m  = \binom{N}{k} \left(\frac{\eta}{2x-1} + 1\right)^{k},
\end{align}
where we used the fact that $\binom{n}{k} =0$ if $k <0$, and the identity $\sum_{k=0}^n \binom{n}{k}\binom{k}{r} x^k = x^r (1+x)^{n-r} \binom{n}{r}$.  Returning to \refew{part2}, we find 
\begin{align}
Z_{N}(V, T, E_0, \Delta) & =\frac{(V/\lambda_0^3)^{2N}}{\sqrt{\pi}} \int^{\infty}_{0} dx\, \frac{e^{-x}}{\sqrt{x}}(2x)^N \sum_{k=0}^N \binom{N}{k} \frac{1}{(2N-2k-1)!!} \left[\frac{\delta}{2x}\left( \eta  + 2x-1\right)\right]^k\mm
& = \frac{(V/\lambda_0^3)^N 2^N}{\sqrt{\pi} (2N-1)!!} \int^{\infty}_{0} dx\, \frac{e^{-x}}{\sqrt{x}}x^N \sum_{k=0}^N \binom{2N}{2k} (2k-1)!! \left[\frac{\delta}{2x}\left( \eta  +  2x-1\right)\right]^k.
\label{eq:part3}
\end{align}
Then, using the integral identity 
\begin{equation}
\sum_{k=0}^N \binom{2N}{2k} (2k-1)!!\,  \Lambda^k= \frac{1}{2\sqrt{\pi}}\int^{\infty}_{0} dy \, \frac{e^{-y}}{\sqrt{y}} \left[ \left(1+ \sqrt{2 \Lambda y}\right)^{2N}+  \left(1-\sqrt{2 \Lambda y}\right)^{2N} \right],
\end{equation}
derived from the integral definition of $(2k-1)!!$ and the binomial theorem, 
\refew{part3} becomes 
\begin{align}
Z_{N}(V, T, E_0, \Delta)   & = \frac{(V/\lambda_0^3)^{2N}2^N}{2\pi (2N-1)!!} \int^{\infty}_{0} \int^{\infty}_{0}  dx\, dy\, \frac{e^{-x}}{\sqrt{x}} \frac{e^{-y}}{\sqrt{y}} \, x^{N} \left[ \left(1 + \sqrt{y{\delta}\left(\eta + 2x-1\right)/x}\right)^{2N} + (\sqrt{y} \to - \sqrt{y}) \right]\mm
&  = \frac{(V/\lambda_0^3)^{2N} 2^N}{2\pi (2N-1)!!} \int^{\infty}_{0}  \int^{\infty}_{0} dx\, dy\, \frac{e^{-x-y}}{\sqrt{xy}} \, \left[ \left(\sqrt{x} + \sqrt{y\delta\left(\eta + 2x-1\right)}\right)^{2N} + \left(\sqrt{y} \to - \sqrt{y}\right) \right],
\end{align}
where $ (\sqrt{y} \to - \sqrt{y})$ stands in for the preceding term with $\sqrt{y}$ replaced with $-\sqrt{y}$. Next, using the identity 
\begin{equation}
 (2N-1)!! = \frac{2^N}{\sqrt{\pi}}  \Gamma\left(N+1/2\right), 
\end{equation}
gives the final form of the partition function.

\section{Equilibrium Conditions for Non-Gendered System \label{app:eqbm_ng}}
In  this section, we justify the conditions defining the Laplace's method approximation of the partition function and show that they result in a system of equations  for $\langle k \rangle$ and $\langle m \rangle$, the average number of dimers and the average number of correct dimers, respectively. 

In the main text, we made the approximation 
\begin{align}
Z_{N}(V, T; E_0, \Delta) & = 
\int^{\infty}_{0} \int^{\infty}_{0} dx \,dy\, \exp\big[ - \beta F_N(x, y;  V, T, E_0,  \Delta) \big] \mm
&  \simeq 2\pi  \left(\det H \right)^{-1/2} \exp\big[ - \beta  F_N(\bar{x}, \bar{y};  V, T, E_0,  \Delta) \big],
\label{eq:approx_part}
\end{align}
where we defined 
\begin{align}
\beta F_N(x, y; & V, T, E_0,  \Delta)  \equiv  x + y + \frac{1}{2} \ln (xy)  - \ln \left({\cal M}_{+}^{2N} + {\cal M}^{2N}_{-} \right) + \beta F_0(N,  V, T), 
\label{eq:sfdef1}
\end{align}
with $\beta F_0(N, V, T)$ composed of terms that are independent of the variables $x$ and $y$ and of the parameters $E_0$ and $\Delta$. In \rfw{approx_part},  $\bar{x}$ and $\bar{y}$ are the critical points of $ F_N(x, y;  V, T, E_0,  \Delta)$, defined by 
\begin{equation}
\partial_{i} (\beta F_N)\Big|_{x = \bar{x}, y = \bar{y} } = 0, 
\label{eq:crit_conds2}
\end{equation}
for $i = x, y$, and $H$ is the Hessian matrix with the elements
\begin{equation}
H_{ij} = \partial_{i} \partial_{j} (\beta F_N)\Big|_{x = \bar{x}, y = \bar{y}}.
\end{equation} 
For the validity of \rfw{approx_part}, $H$ must satisfy
\begin{equation}
\det H > 0\,, \quad \Tr H>0.
\label{eq:hessian2}
\end{equation}
Equivalently, \rfw{hessian2} also ensures that the critical points defined by \rfw{crit_conds} are stable. We can compute the average number of dimers and the average number of correct dimers from the partition function via 
\begin{equation}
\langle k \rangle = \frac{\partial}{\partial(\beta E_0)} \ln Z_{N}(V, T; E_0, \Delta), \qquad \langle m \rangle = \frac{\partial}{\partial (\beta \Delta)} \ln Z_{N}(V, T; E_0, \Delta).
\label{eq:kandmdef}
\end{equation}

In Sec. \ref{app:crits}, we will use the conditions \refew{crit_conds2} along with the definitions in \refew{kandmdef} to calculate equilibrium constraints on $\langle k \rangle$ and $\langle m \rangle$. In Sec. \ref{app:stabs}, we will show the equilibria derived from these conditions satisfy \rfw{hessian2} and are indeed stable. Also, by computing the Hessian, we will show that the $\ln \det H$ contribution the Hessian could make to the free energy in \rfw{sfdef1} is sub-leading in the large $N$ limit because it is of the same order as the terms we drop in our derivation of the equilibrium conditions.

\subsection{Computing Critical Points \label{app:crits}}
Here we will derive the equilibrium conditions on $\langle k \rangle$ and $\langle m \rangle$ resulting from a $N\gg1$ approximation of the partition function. We write the free energy \refew{sfdef1} slightly differently as  
\begin{equation}
\beta F_N(x, y;  V, T, E_0,  \Delta) = x + y + (1/2- N) \ln x + \frac{1}{2} \ln y - \ln \left({\cal N}_{+}^{2N} + {\cal N}^{2N}_{-} \right) + \beta F_0(N, V, T), 
\label{eq:fdefa}
\end{equation}
where 
\begin{equation}
{\cal N}_{\pm} \equiv 1 \pm \delta^{1/2} \sqrt{y\, \Lambda(x; \beta \Delta)}, 
\label{eq:Ndef}
\end{equation}
and 
\begin{equation}
\Lambda(x; \beta \Delta) \equiv \frac{e^{\beta \Delta}-1}{x} + 2\,, \qquad \delta \equiv \frac{2 \sqrt{2}\lambda_0^3}{V} e^{\beta E_0}.
\label{eq:Lamdef}
\end{equation}
We can simplify \refew{fdefa} by considering our presumed $N\gg1$ limit. First we note that $(1+ Q)^N + (1- Q)^N =  (1+Q)^N(1 + \phi_N)$ where,  if  $Q >0$, then $\phi_N \to 0$ for $N \to \infty$. Thus, \refew{fdefa} can be written as 
\begin{equation}
\beta F_N(x, y;  V, T, E_0,  \Delta) =x + y - N \ln x + \frac{1}{2} \ln y -  2N \ln {\cal N}_{+}+ \beta F_{0}(N, V, T) + {\epsilon}_N, 
\label{eq:fdef3}
\end{equation}
where $\epsilon_N$ is the error term which includes all terms that are subleading in the $N\gg 1$ limit. Now, using \refew{Ndef} and \refew{fdef3}, we see that \refew{crit_conds2} yields the equations 
\begin{align}
0 & = \partial_{x} (\beta F_N) \Big|_{x = \bar{x}, y = \bar{y} }  = 1 - \frac{N}{\bar{x}}  - \frac{N\delta^{1/2} \sqrt{\bar{y}/\Lambda(\bar{x}; \beta \Delta)}}{1 + \delta^{1/2} \sqrt{\bar{y} \Lambda(\bar{x}; \beta \Delta)}}\cdot \left( - \frac{e^{\beta \Delta}-1}{\bar{x}^2}\right), \label{eq:xcrit}\\
0 & = \partial_{y} (\beta F_N) \Big|_{x = \bar{x}, y = \bar{y} }  = 1 + \frac{1}{2\bar{y}} - \frac{N\delta^{1/2} \sqrt{\Lambda(\bar{x}; \beta \Delta)/\bar{y}}}{1 + \delta^{1/2} \sqrt{\bar{y} \Lambda(\bar{x}; \beta \Delta)}}.
\label{eq:ycrit}
\end{align}
From the definitions in \refew{kandmdef}, we can express $\langle k \rangle$ and $\langle m \rangle$ in terms of $\bar{x}$ and $\bar{y}$: 
\begin{align}
\langle k \rangle &=  \partial_{\beta E_0} \ln Z_{N} = - \partial_{\beta E_0} (\beta F_{N}) \Big|_{x = \bar{x}, y = \bar{y} }\mm
& = \frac{N \delta^{1/2} \sqrt{\bar{y} \Lambda(\bar{x}; \beta \Delta)}}{1 + \delta^{1/2} \sqrt{\bar{y} \Lambda(\bar{x}; \beta \Delta)}} \label{eq:kcrit} \\
\langle m \rangle &=  \partial_{\beta \Delta} \ln Z_{N} = - \partial_{\beta \Delta} (\beta F_{N}) \Big|_{x = \bar{x}, y = \bar{y} }\mm
& = \frac{N \delta^{1/2} \sqrt{\bar{y} /\Lambda(\bar{x}; \beta \Delta)}}{1 + \delta^{1/2} \sqrt{\bar{y} \Lambda(\bar{x}; \beta \Delta)}} \cdot \frac{e^{\beta \Delta}}{\bar{x}} \label{eq:mcrit},
\end{align}
where we used \refew{xcrit} and \refew{ycrit} to set the coefficients of $\partial \bar{x}/ \partial(\beta E_0)$ and $\partial \bar{y}/ \partial(\beta E_0)$ (and similarly for the $\bar{x}$ and $\bar{y}$ derivatives with respect to $\beta \Delta$) to zero. To be explicit, we note that the second equalities in both \rfw{kcrit} and \rfw{mcrit} would be better expressed as approximations derived from \rfw{approx_part}. However, for the analytical calculations of this system we will always be working in the $N\gg1$ regime and we will take the free energy \rfw{fdef3} as the true free energy of the system. 

From \refew{ycrit}, we find the condition
\begin{equation}
\bar{y} + {1}/{2} = \frac{N \delta^{1/2} \sqrt{\bar{y} \Lambda(\bar{x}; \beta \Delta)}}{1 + \delta^{1/2} \sqrt{\bar{y} \Lambda(\bar{x}; \beta \Delta)}},
\label{eq:ycrit1}
\end{equation}
and with \refew{kcrit}, we obtain 
\begin{equation}
\bar{y}  + 1/2 =  \langle k \rangle.
\label{eq:ykdef}
\end{equation}
Inverting \refew{ycrit1}, we find 
\begin{equation}
 \delta\, \bar{y} \,\Lambda(\bar{x}; \beta \Delta) = \frac{(\bar{y} + 1/2)^2}{\left( N - (\bar{y} + 1/2) \right)^2} , 
\end{equation}
or, with \refew{ykdef}, 
\begin{equation}
\delta\, \big(\langle k \rangle - 1/2\big) \,\Lambda(\bar{x}; \beta \Delta) = \frac{\langle k \rangle^2}{\big( N - \langle k \rangle \big)^2} .
\label{eq:kcrit1}
\end{equation}
We can further reduce this result by solving for $\Lambda(\bar{x}; \beta \Delta)$ in terms of $\langle k \rangle$ and $\langle m \rangle$. Dividing \refew{kcrit} by \refew{mcrit}, yields 
\begin{equation}
\frac{\langle k \rangle}{\langle m \rangle} = \bar{x}\, \Lambda(\bar{x}; \beta \Delta) e^{-\beta \Delta},
\label{eq:km} 
\end{equation}
which when solved for $\bar{x}$, gives us
\begin{equation}
\bar{x} = \frac{1}{2}\left[ 1+ \frac{\langle k \rangle - \langle m \rangle}{\langle m \rangle} e^{\beta \Delta}\right]
\label{eq:xcrit1}.
\end{equation}
Substituting \refew{xcrit1} into \refew{km}, then gives us 
\begin{equation}
\Lambda(\bar{x}; \beta \Delta) = \frac{2 \langle k \rangle }{\langle k \rangle - \langle m \rangle (1 - e^{-\beta \Delta}) }. 
\label{eq:km2}
\end{equation}
Returning to \refew{kcrit1}, we obtain 
\begin{equation}
2 \delta \left(1 - \frac{1}{2\langle k \rangle}\right) = \frac{\langle k \rangle - \langle m \rangle (1 - e^{-\beta \Delta})}{\big(N- \langle k \rangle\big)^2 }.
\label{eq:kmeqbm01}
\end{equation}
which is the first equilibrium condition constraining $\langle k \rangle$ and $\langle m \rangle$. We will primarily be interested in temperature ranges at which $\langle k \rangle$ assumes a non-trivial value much larger than of ${\cal O}(1)$. Thus we can take $\langle k \rangle \gg 1$ leading to the result 
\begin{equation}
\frac{4 \sqrt{2} \lambda_0^3}{V}\, e^{\beta E_0} = \frac{\langle k \rangle - \langle m \rangle (1 - e^{-\beta \Delta})}{\big(N- \langle k \rangle\big)^2 }+ {\cal O}(\langle k \rangle ^{-1})
\label{eq:kmeqbm1}
\end{equation}

To find the second equilibrium condition, we note that \refew{xcrit} can be written as 
\begin{equation}
(N- \bar{x}) \Lambda(\bar{x}; \beta \Delta ) \bar{x} 
 = \langle k \rangle (e^{\beta \Delta}-1). 
\end{equation}
Using \refew{km} and \refew{xcrit1}, this result becomes 
\begin{equation}
N - \frac{1}{2}\left[ 1 + \frac{\langle k \rangle - \langle m \rangle}{\langle m \rangle} e^{\beta \Delta} \right] = \langle m \rangle (1- e^{-\beta \Delta}),
\end{equation}
or, with some rearranging, 
\begin{equation}
\frac{e^{\beta \Delta}}{2} = \langle m \rangle  \frac{N - \langle m \rangle (1- e^{-\beta \Delta})}{\langle k \rangle - \langle m \rangle (1- e^{-\beta \Delta})}, 
\label{eq:kmeqbm2}
\end{equation}
which is our second equilibrium condition.  With the equilibrium conditions \refew{kmeqbm1} and \refew{kmeqbm2} established, we can now turn to showing that these equilibria define stable minima of the free energy.  

\subsection{Demonstrating Stability \label{app:stabs}}
To check whether the equilibrium conditions \refew{kmeqbm1} and \refew{kmeqbm2} define stable equilibria for this system, we need to compute the various elements of the Hessian matrix
\begin{equation}
H_{ij} = \partial_{i} \partial_{j} (\beta F_N)\Big|_{x = \bar{x}, y = \bar{y} },
\end{equation} 
and ensure that the matrix is positive definite. By definition, a positive definite matrix is one with positive eigenvalues. For the $2\times 2$ matrix considered here, this amounts to having a positive determinant and positive trace:
\begin{equation}
 \Tr H >0, \qquad  \det H >0.
\end{equation}

We will first compute the diagonal elements composing $\Tr H$. To compute $\partial_{y}^2(\beta F_N)|_{x = \bar{x}, y = \bar{y}}$, we must compute the first and second-order $y$ derivatives of the free energy as general functions. Given \rfw{fdef3}, we obtain 
\begin{align}
\partial_{y}(\beta F_N) & = 1 + \frac{1}{2y} - \frac{2N}{{\cal N}_{+}} \partial_{y} {\cal N}_+\label{eq:p1yF} \\
\partial_{y}^2(\beta F_N) & = -\frac{1}{2y^2}  + 2N \left[ \frac{(\partial_{y}{\cal N})_+^2}{{\cal N}_+^2} - \frac{\partial_{y}^2 {\cal N}_{+}}{{\cal N}_+}  \right].
\label{eq:p2yF0}
\end{align}
From \refew{Ndef}, we have 
\begin{equation}
 \partial_{y} {\cal N}_+ =  \frac{\delta^{1/2}}{2} \sqrt{\frac{\Lambda(x; \beta \Delta)}{y}}, \qquad \partial_{y}^2 {\cal N}_+ =  -\frac{\delta^{1/2}}{4} \sqrt{\frac{\Lambda(x; \beta \Delta)}{y^3}} = - \frac{1}{2y}  \partial_{y} {\cal N}_+.
\end{equation}
Thus, \refew{p2yF0} becomes 
\begin{align}
\partial_{y}^2(\beta F_N) & = -\frac{1}{2y^2}  + 2N \left[ \frac{(\partial_{y}{\cal N})_+^2}{{\cal N}_+^2} + \frac{1}{2y} \frac{\partial_{y} {\cal N}_{+}}{{\cal N}_+}  \right].
\label{eq:p2yF1}
\end{align}
Setting $x = \bar{x}$ and $y = \bar{y}$ in \refew{p2yF1} and noting that $\partial_{y}(\beta F_N) = 0$ at these values, we find 
\begin{align}
\partial_{y}^2(\beta F_N) \Big|_{x = \bar{x}, y = \bar{y}}  
 & = \frac{1}{2N \bar{y}^2} \big[- N +  (\bar{y} + 1/2)^2 + N (\bar{y} + 1/2)  \big],
 \label{eq:p2yF}
\end{align}
where we used \refew{p1yF} evaluated at $x = \bar{x}$ and $y = \bar{y}$. 
Considering the argument of the above expression, we find that it is positive for $\bar{y} > 1/2 + {\cal O}(N^{-1})$. In terms of our order parameter,  this result translates into $\partial_{y}^2(\beta F_N)|_{x = \bar{x}, y = \bar{y}}$ being positive for $\langle k \rangle >1$ which is only violated when we are well-outside the range for non-trivial values of $\langle k \rangle$. 

Next, computing $\partial_{x}^2 (\beta F_N)$ given \rfw{fdef3}, we obtain
\begin{align}
\partial_{x}(\beta F_N) & = 1 - \frac{N}{x} - \frac{2N}{{\cal N}_{+}} \partial_{x} {\cal N}_+\label{eq:p1xF} \\
\partial_{x}^2(\beta F_N) & = \frac{N}{x^2}  + 2N \left[ \frac{(\partial_{x}{\cal N})_+^2}{{\cal N}_+^2} - \frac{\partial_{x}^2 {\cal N}_{+}}{{\cal N}_+}  \right], 
\label{eq:p2xF}
\end{align}
where 
\begin{equation}
\partial_{x} {\cal N}_{+} = \frac{\delta^{1/2}}{2}\sqrt{\frac{y}{\Lambda(x; \beta \Delta)}}\cdot \partial_{x}\Lambda(x; \beta \Delta), 
\label{eq:pxNp}
\end{equation}
and 
\begin{align}
\partial^2_x{\cal N}_+ 
& = \frac{\partial_{x} {\cal N}_{+} }{\partial_{x}\Lambda(x; \beta \Delta)} \cdot \frac{1}{\Lambda(x; \beta \Delta)}\cdot \left[\Lambda(x; \beta \Delta)\,\partial_{x}^2 \Lambda(x; \beta \Delta) - \frac{1}{2}\left(\partial_{x}\Lambda(x; \beta\Delta)\right)^2  \right]. 
\label{eq:p2Nx}
\end{align}
Using the definition of $\Lambda(x; \beta \Delta)$ (given in \refew{Lamdef}) in the quantity in the brackets above yields
 \begin{align}
  \left[\Lambda(x; \beta \Delta)\,\partial_{x}^2 \Lambda(x; \beta \Delta) - \frac{1}{2}\left(\partial_{x}\Lambda(x; \beta\Delta)\right)^2  \right] 
  & = - \frac{\partial_{x}\Lambda(x; \beta \Delta)}{x} \left[ \frac{3}{2} \Lambda(x; \beta \Delta) + 1\right].
 \end{align}
Thus \refew{p2Nx} becomes 
\begin{equation}
\partial^2_x{\cal N}_+ = - \frac{\partial_x {\cal N}_+}{x} \left[ \frac{3}{2} +\frac{1}{\Lambda(x; \beta \Delta)}\right].
\end{equation}
Now, returning to \refew{p2xF} we have 
\begin{equation}
\partial_{x}^2(\beta F_N) = \frac{N}{x^2} + 2N\left(\frac{\partial_x {\cal N}_{+}}{{\cal N}_+}\right)^{2}  + 2N \frac{\partial_x {\cal N}_{+}}{{\cal N}_+} \left[\frac{3}{2x} + \frac{1}{x\Lambda(x; \beta \Delta)}\right].
\label{eq:p2xFs}
\end{equation}
Setting $x = \bar{x}$ and $y = \bar{y}$ in \refew{p2xFs} and noting that $\partial_{x}(\beta F_N) = 0$ at these values, we obtain 
\begin{align}
\partial_{x}^2(\beta F_N)  \Big|_{x = \bar{x} , y = \bar{y}} 
& = \frac{1}{2N \bar{x}^2} \left[\bar{x} (\bar{x} + N) + \frac{2 N (\bar{x}- N)}{\Lambda(\bar{x}; \beta \Delta)} \right].
\label{eq:p2xF0} 
\end{align}
We can make further progress by expressing $\Lambda(\bar{x}; \beta \Delta)$ in terms of $\bar{x}$ and $\bar{y}$. First, we note that \refew{xcrit1} and \refew{kmeqbm2} together yield
\begin{equation}
\bar{x} = N - \langle m \rangle (1- e^{-\beta \Delta}), 
\label{eq:xNm}
\end{equation}
and inverting \refew{km2} gives us 
\begin{equation}
\frac{1}{\Lambda(\bar{x};\beta\Delta)} = \frac{1}{2} \left(1 - \frac{\langle m \rangle (1- e^{-\beta \Delta}) }{\langle k \rangle} \right) = \frac{1}{2} \left( 1 - \frac{N-\bar{x}}{\bar{y} + 1/2}\right),
\end{equation}
where we used \refew{xNm} and \refew{ykdef} in the final equality. Returning to \refew{p2xF0}, we find
\begin{align}
\partial_{x}^2(\beta F_N)  \Big|_{x = \bar{x} , y = \bar{y}} 
& = \frac{1}{2N \bar{x}^2} \left[ \bar{x}^2 \left( \lambda +1\right) - 2N \bar{x} \left( \lambda -1\right) + N^2 \left(\lambda - 1\right) \right],
\label{eq:p2xF1} 
\end{align}
where we defined 
\begin{equation}
\lambda \equiv \frac{N}{\bar{y}+1/2}. 
\label{eq:lamydef}
\end{equation}
Since $\bar{y} + 1/2 = \langle k \rangle$ and $\langle k \rangle <N$, we have $\lambda > 1$ for non-zero temperature. For the function 
\begin{equation}
f(z) = z^2 (\lambda+1) - 2 z ( \lambda-1) + \lambda-1
\end{equation}
where $z \in \mathbb{R}^{+}$, it can be shown that the minimum satisfies
\begin{equation}
\left[ f(z) \right]_{\text{min}} = \frac{\lambda-1}{\lambda+1}. 
\end{equation}
Thus, for $\lambda>1$,  we find that $f(z) >0$. Therefore, \refew{p2xF1} is greater than zero for equilibrium values $\bar{x}$ and $\bar{y}$. With \refew{p2yF} and \refew{p2xF1}, we can thus conclude
\begin{equation}
\Tr H = \left[ \partial_{y}^2(\beta F_N)   + \partial_{x}^2(\beta F_N)  \right]\Big|_{x = \bar{x} , y = \bar{y}} > 0, 
 \label{eq:trH}
\end{equation}
for $\langle m \rangle$ and $\langle k \rangle$ constrained by \refew{kmeqbm1} and \refew{kmeqbm2}. 

Now, we compute the off-diagonal elements that make up (together with the diagonal elements) $\det H$. Taking the $y$-partial derivative of \refew{p1xF}, we have
\begin{equation}
\partial_{y} \partial_{x} (\beta F_N) = -2N\left[ \frac{1}{{\cal N}_+} \partial_{y} \partial_{x} {\cal N}_+ - \frac{1}{{\cal N}_+^2} \partial_{x} {\cal N}_+ \partial_{y} {\cal N}_{+} \right].
\label{eq:pxpyF}
\end{equation}
From \refew{Ndef}, we have that the mixed partial of ${\cal N}_+$ is 
\begin{align}
\partial_{y} \partial_x {\cal N}_{+} &= \partial_{y} \left[ \frac{\delta^{1/2}}{2}\sqrt{\frac{y}{\Lambda(x; \beta \Delta)}}\cdot \partial_{x}\Lambda(x; \beta \Delta)\right]\mm
& = \frac{\delta^{1/2}}{4}\sqrt{\frac{1}{y\Lambda(x; \beta \Delta)}}\cdot \partial_{x}\Lambda(x; \beta \Delta) =  \frac{1}{2y} \partial_x {\cal N}_{+},
\end{align}
where we used \refew{pxNp}, in the final equality. 
 Evaluating \refew{pxpyF} at $x = \bar{x}$ and $y = \bar{y}$ and using 
 \begin{equation}
\frac{1}{{\cal N}_+} \partial_{y} {\cal N}_{+}\Big|_{x = \bar{x}, y = \bar{y}} = \frac{1}{2N} \left( 1+ \frac{1}{2\bar{y}}\right), \qquad \frac{1}{{\cal N}_+} \partial_{x} {\cal N}_{+}\Big|_{x = \bar{x}, y = \bar{y}} = \frac{1}{2N} \left( 1- \frac{N}{\bar{x}}\right), 
\end{equation}
found from \rfw{p1yF}, \rfw{p1xF}, and the critical point condition, we obtain 
\begin{align}
\partial_{y} \partial_{x} (\beta F_N) \Big|_{x = \bar{x}, y = \bar{y}}  
& = \frac{(N- \bar{x})\big(N - \bar{y}-1/2 \big)}{2N \bar{x} \bar{y}}.
\label{eq:pypxF}
\end{align}
Before we compute the determinant, it will prove useful to express the $\bar{y}$ in \refew{p2yF} and \refew{pypxF} in terms of $\lambda$ given in \rfw{lamydef}. From \refew{lamydef}, we find 
\begin{align}
\partial_{y}^2 (\beta F_N) \Big|_{x = \bar{x}, y = \bar{y}}  & = \frac{(\bar{y} + 1/2)^2}{2N \bar{y}^2}\left( - \frac{N}{(\bar{y} + 1/2)^2} + 1 +\frac{N}{\bar{y}+1/2}\right) \mm
 & = \frac{N}{2\lambda^2 \bar{y}^2}\left( - \frac{\lambda^2}{N} + 1 + \lambda\right) \label{eq:p2yFlam} \\
 \partial_{y} \partial_{x} (\beta F_N) \Big|_{x = \bar{x}, y = \bar{y}}   & = \frac{1}{2 N \bar{x} \bar{y}}(\bar{y}+1/2)(N-\bar{x}) \left( \frac{N}{\bar{y}+1/2}- 1\right)\mm
  & = \frac{1}{2 \lambda \bar{x} \bar{y}}(N-\bar{x}) \left( \lambda- 1\right) \label{eq:pypxFlam}
\end{align}
Finally, computing the determinant of the Hessian from \rfw{p2xF1}, \rfw{p2yFlam}, and \rfw{pypxFlam}, we thus find 
\begin{align}
\det H&  = \left[ \partial_{y}^2(\beta F_N)\partial_{x}^2(\beta F_N) -   \left(\partial_{y} \partial_{x} (\beta F_N) \right)^2  \right]\Big|_{x = \bar{x} , y = \bar{y}} \mm
 & = \frac{1}{4 \lambda^2 \bar{x}^2 \bar{y}^2} \left[A_{\lambda} \bar{x}^2 - 2N B_{\lambda} \bar{x} + B_{\lambda}\right],
 \label{eq:detH}
\end{align}
where 
\begin{align}
A_{\lambda} & 
\equiv \lambda\left( 4 - \frac{\lambda(\lambda+1)}{N}\right)\label{eq:Alam} \\
 B_{\lambda} & 
\equiv (\lambda-1)\left(2 - \frac{\lambda^2}{N}\right) \label{eq:Blam}.
\end{align}
We want to show that \refew{detH} is always positive. We will employ a method similar to that used in showing that $\partial_{x}^2(\beta F_N)|_{x = \bar{x}, y = \bar{y}}$ is positive. For the function 
\begin{equation}
g(z) = A_{\lambda} z^2 - 2 B_{\lambda} z + B_{\lambda},
\end{equation}
where $z \in \mathbb{R}^{+}$, it can be shown that the minimum is given by
\begin{equation}
[g(z)]_{\text{min}} = B_{\lambda} \left( 1  - \frac{B_{\lambda}}{A_{\lambda}}\right). 
\label{eq:gmin}
\end{equation}
From \refew{Alam}, \refew{Blam}, and the condition $1 < \lambda< N$, we find that $ B_{\lambda} < A_{\lambda} $ for all valid $\lambda$. From this inequality, we find 
\begin{equation}
 \frac{B_{\lambda}}{A_{\lambda}} < 1, \, \, \text{ if $B_{\lambda} >0$,} \quad \text{ and } \quad \frac{B_{\lambda}}{A_{\lambda}}> 1, \, \, \text{ if $B_{\lambda} <0$.}
\end{equation}
 Thus, we can conclude that \refew{gmin} is always positive for the entire domain of $z$ and for valid values of $\lambda$. Considering \refew{detH} we then have 
\begin{equation}
\det H = \left[ \partial_{y}^2(\beta F_N)\partial_{x}^2(\beta F_N) -   \left(\partial_{y} \partial_{x} (\beta F_N) \right)^2  \right]\Big|_{x = \bar{x} , y = \bar{y}} > 0. 
 \label{eq:detH1}
\end{equation}
With \refew{trH} and \refew{detH1}, we can conclude that the Hessian matrix is positive definite and thus that the derived equilibrium conditions \refew{kmeqbm1} and \refew{kmeqbm2} define stable equilibria of the free energy \refew{fdef1}, and, moreover, that our Laplace's method approximation of the partition function \refew{fapprox} is valid.

Finally, in \rfw{detH}, we see that $\ln \det H$ is on the order of a linear combination of $\ln \bar{x}$ and $\ln \bar{y}$. Given that we ultimately dropped such terms from our calculation of the equilibrium conditions \rfw{kmeqbm1} and \rfw{kmeqbm2}, we now see that we were also justified in ignoring the $\ln \det H$ contributions to our free energy.

\section{Simulation of Dimer System \label{app:mcmc}}

The simulation results in \reffig{typeIandII} were obtained using the Metropolis-Hastings Monte Carlo algorithm. We defined the microstate of our system by two lists: One defining the particles that are monomers and the other defining the monomer-monomer pairs making up the dimers. For example, a $2N = 10$ particle system could have a microstate defined by the monomer list $[1,4,6,9]$ and the dimer list $[(3,5), (2,8), (7, 10)]$. The free energy of a microstate was given by 
\begin{equation}
f(k, m) = - k E_0 - m \Delta - k\,k_BT\ln (V/\lambda_0^3) - (2N-2k)k_BT \ln(2 \sqrt{2}\, V/\lambda_0^3), 
\label{eq:micro_freeen}
\end{equation}
for a system with $k$ dimers of which $m$ consisted of correct dimers. 

To efficiently explore the state space of the system, we used three different types of transitions with unique probability weights for each one. In the following, $N_{\text{m}}$ and $N_{\text{d}}$ represent the lengths of the monomer and dimer lists, respectively, before the transition. 
\begin{enumerate}
\item \tbf{Monomer Association:} Two randomly chosen monomers are removed from the monomer list, joined as a pair, and the pair is appended to the dimer list.  Weight = $\binom{N_{\text{m}}}{2}/(N_{\text{d}}+1)$\\ \textit{Example:} mon = $[1,3, 4, 5, 6,9]$ and dim = $[(2,8), (7, 10)]$ $\to$ mon = $[1,4,6,9]$ and\\ dim = $[(3,5), (2,8), (7, 10)]$; Weight = 15/3.

\item \tbf{Dimer Dissociation:} One randomly chosen dimer is removed from the dimer list, and both of its elements are appended to the monomer list. Weight = $ N_{\text{d}}/\binom{N_{\text{m}}+2}{2}$

\textit{Example:} mon = $[6,9]$ and dim = $[ (1, 4), (3,5), (2,8), (7, 10)]$ $\to$ mon = $[2, 6,8, 9]$ and\\ dim = $[ (1, 4), (3,5), (7, 10)]$; Weight = 4/6.

\item \tbf{Dimer Cross-Over:} Two dimers are chosen randomly. One randomly chosen element from one dimer is switched with a randomly chosen element of the other dimer.  Weight =1. 

\textit{Example:} dim = $[ (1, 4), (3,5) (7, 10) ]$ $\to$ dim = $[ (1, 10), (3,5), (4, 10)]$  ]; Weight = 1.

\end{enumerate}
The third type of transition is unphysical but is necessary to ensure that the system can quickly escape kinetic traps that lead to inefficient sampling of the state space. 

For each simulation step, there was a 1/3 probability of selecting a particular transition type and the suggested transition was accepted with log-probability 
\begin{equation}
\ln p_{\text{accept}} = -  (f_{\text{fin.}}-f_{\text{init.}})/k_BT+\ln\text{(Weight)}, 
\label{eq:pacc}
\end{equation}
where $f_{\text{fin}}$ and $f_{\text{init}}$ are the final and initial free energies of the microstate defined according to \rfw{micro_freeen}, and (Weight) is the ratio between the number of ways to make the forward transition and the number of ways to make the reverse transition. This weight was chosen for each transition type to ensure that detailed balance was maintained. For impossible transitions (e.g., monomer association for a microstate with no monomers), $p_{\text{accept}}$ was set to zero. 

At each temperature, the simulation was run for $30,000$ time steps, of which the last 600 were used to compute ensemble averages of $\langle k \rangle$ and $\langle m \rangle$. These simulations were repeated 50 times, and each point in \reffig{typeIandII} represents the average $\langle k \rangle$ and $\langle m \rangle$ over these runs. Jupyter notebooks for these simulations are linked to in the \textit{Supplementary Code}. 

\section{Temperature Changes in Parameter Space \label{app:tempchange}}
\begin{figure}[t]
\begin{centering}
\begin{subfigure}{0.3\textwidth}
\centering
\includegraphics[width=\linewidth]{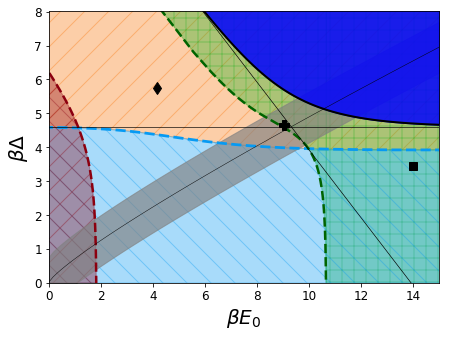}
	\caption{}
	\label{fig:phs1_temp1}
\end{subfigure} 
\begin{subfigure}{0.3\textwidth}
\centering
\includegraphics[width=\linewidth]{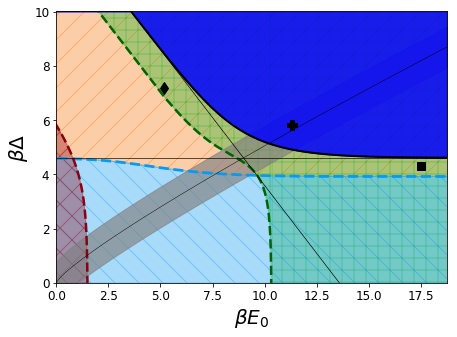}
	\caption{}
	\label{fig:phs1_temp2}
\end{subfigure} 
\begin{subfigure}{0.3\textwidth}
  \centering
  \includegraphics[width=\linewidth]{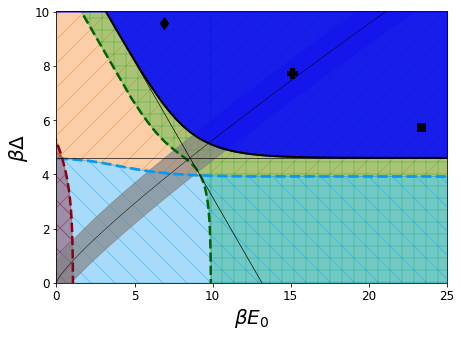}
	\caption{}
	\label{fig:phs1_temp3}
\end{subfigure}
\begin{subfigure}{0.3\textwidth}
\centering
\includegraphics[width=\linewidth]{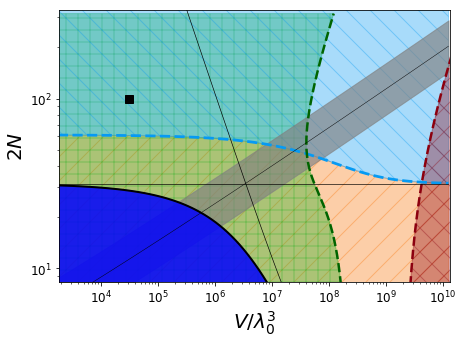}
	\caption{}
	\label{fig:phs2_temp1}
\end{subfigure} 
\begin{subfigure}{0.3\textwidth}
\centering
\includegraphics[width=\linewidth]{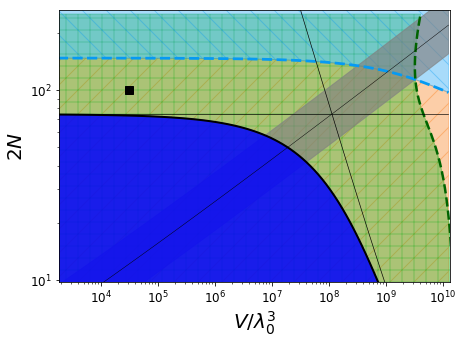}
	\caption{}
	\label{fig:phs2_temp2}
\end{subfigure} 
\begin{subfigure}{0.3\textwidth}
  \centering
  \includegraphics[width=\linewidth]{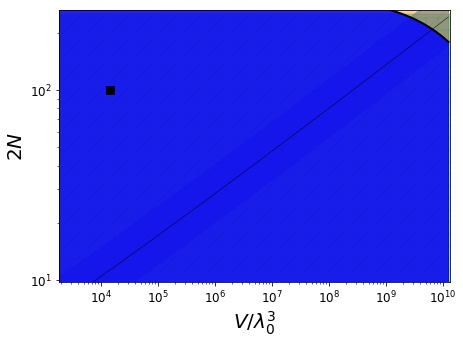}
	\caption{}
	\label{fig:phs2_temp3}
\end{subfigure}
	\caption{(a), (b), and (c): Plots of the points in Fig. 5a in the main text as we lower the system temperature: (a) depicts $k_BT=1.0$ (i.e., the same temperature as the original figure), (b) depicts $k_BT = 0.8$, and (c) depicts $k_BT = 0.6$. Consistent with the simulation plots in Fig. 4, at $k_BT = 0.6$ all the systems are in the "fully correct dimerization" regime. (d), (e), and (f): Plots of the point in Fig. 5b as we lower the system temperature: (d) depicts $k_BT=1.0$ (i.e., the same temperature as the original figure), (e) depicts $k_BT = 0.8$, and (f) depicts $k_BT = 0.6$. Consistent with the simulation plots in Fig. 4(c), at $k_BT = 0.6$ the systems is in the "fully correct dimerization" regime.}
	\label{fig:phaseplots_temp}
\end{centering}
\end{figure}

In \reffig{phaseplots_temp} we depict how the plots in Fig. 5 change as we change the value of $k_BT$. We note that since the regions are defined by temperature dependent boundaries, changing the temperature of a system represented by a point also changes the arrangement of the boundaries that surround the point.

\bibliographystyle{ieeetr}

\end{document}